\documentclass[11pt, a4paper]{article}
\pdfoutput=1
\usepackage{jcappub}
\usepackage{graphicx}
\usepackage{graphics}
\usepackage{dcolumn}
\usepackage{bm}		
\usepackage{tabularx}
\usepackage{slashed}
\usepackage{verbatim}
\usepackage{lipsum}
\usepackage{braket}
\usepackage{physics}
\usepackage{float}
\usepackage{natbib}
\usepackage{color}
\usepackage[dvipsnames]{xcolor}
\usepackage{xcolor}
\usepackage{xspace}
\usepackage{hhline}
\usepackage{multirow}
\usepackage{amsmath}
\usepackage{amssymb}
\usepackage{xspace}
\usepackage{textcomp}
\usepackage{gensymb}
\usepackage{booktabs}
\usepackage{hyperref}
\usepackage{cleveref}
\usepackage{soul}
\usepackage{array}
\usepackage{adjustbox}
\usepackage{cancel}
\usepackage{fancyhdr}
\usepackage{aas_macros}

\newcommand{\beq}{\begin{equation}}
\newcommand{\baln}{\begin{aligned}}
\newcommand{\eeq}{\end{equation}}
\newcommand{\ealn}{\end{aligned}}
\newcommand{\beqn}{\begin{eqnarray}}
\newcommand{\eeqn}{\end{eqnarray}}
\newcommand{\burst}{{\sc burst}\xspace}
\newcommand{\rhodp}{\ensuremath{\rho_{A^{\prime}}}\xspace}
\newcommand{\taudp}{\ensuremath{\tau_{A^{\prime}}}\xspace}

\newcommand{\massdp}{\ensuremath{m_{A^{\prime}}}\xspace}
\newcommand{\spl}{\ensuremath{s_{\rm pl}}\xspace}
\newcommand{\snu}{\ensuremath{s_{\nu}}\xspace}
\newcommand{\tcm}{\ensuremath{T_{\rm cm}}\xspace}
\newcommand{\neff}{\ensuremath{N_{\rm eff}}\xspace}
\newcommand{\dtoh}{\ensuremath{{\rm D/H}}\xspace}
\newcommand{\yp}{\ensuremath{Y_{\rm P}}\xspace}
\newcommand{\feq}{\ensuremath{f^\textnormal{(eq)}}\xspace}

\count\footins = 1000
\title{Probing dark photons in the early universe with big bang nucleosynthesis}
\author[a]{Jung-Tsung Li,}
\author[a]{George M. Fuller}
\author[b]{and Evan Grohs}
\affiliation[a]{Department of Physics, University of California, San Diego, CA 92093, USA}
\affiliation[b]{Department of Physics, University of California, Berkeley, CA 94720, USA}
\emailAdd{jul171@ucsd.edu}
\emailAdd{gfuller@physics.ucsd.edu}
\emailAdd{egrohs@berkeley.edu}

\abstract{We perform calculations of dark photon production and decay in the early universe for ranges of dark photon masses and vacuum coupling with standard model photons. Simultaneously and self-consistently with dark photon production and decay, our calculations include a complete treatment of weak decoupling and big bang nucleosynthesis (BBN) physics. These calculations incorporate all relevant weak, electromagnetic, and strong nuclear reactions, including charge-changing (isospin-changing) lepton capture and decay processes. They reveal a rich interplay of dark photon production, decay, and associated out-of-equilibrium transport of entropy into the decoupling neutrino seas. Most importantly, the self-consistent nature of our simulations allows us to capture the magnitude and phasing of entropy injection and dilution. Entropy injection-induced alteration of the time-temperature-scale factor relation during weak decoupling and BBN leads to changes in the light element abundance yields and the total radiation content (as parametrized by $\neff$). These changes suggest ways to extend previous dark photon BBN constraints. However, our calculations also identify ranges of dark photon mass and couplings not yet constrained, but perhaps accessible and probable, in future Stage-4 cosmic microwave background experiments and future high precision primordial deuterium abundance measurements.}

\keywords{dark photon, big bang nucleosynthesis, beyond standard model, Stage-4 cosmic  microwave  background experiment}

\begin{document}

\maketitle
\flushbottom

\section{Introduction}\label{sec: introduction}

The early universe and the increasingly sophisticated observations that constrain its history together comprise a promising ``laboratory'' for dark sector physics. Vetting new physics with this laboratory demands accurate modeling of the effects of this physics on the evolution of key parameters in the early universe and the impact of those effects on observables. In this paper we examine in a self-consistent way how dark photons in a specific range of masses and couplings with the standard model affect neutrino decoupling, associated relic energy density, and light element abundances.

Weak interaction decoupling and big bang nucleosynthesis (BBN) in the early universe are protracted and intertwined processes. Together they proceed over many Hubble times, roughly spanning temperature regimes from $T \sim 10\,{\rm MeV}$ to $T \sim 10\,{\rm keV}$. In strictly standard model cosmology, neutrino charged- and neutral-current scattering on electrons and positrons continues to facilitate energy and entropy transfer to the plasma of electrons, positrons, photons, and nucleons even down to temperatures near alpha particle formation ($T\sim 100\,{\rm keV}$), although the effectiveness of this transfer decreases significantly with falling temperature. Likewise, charged current weak processes involving neutrinos and charged leptons continue alteration of the neutron-to-proton ($n/p$) ratio throughout this epoch~\cite{Grohs:2016vef}.

If there are out-of-equilibrium Beyond Standard Model (BSM) particles decaying and injecting energy and entropy into the plasma during the weak decoupling epoch, then there will be an extra (over and above the standard model) entropy flow between the neutrino and plasma sectors. This effect not only alters the $n/p$ ratio from standard model cosmology, but also changes the phasing of these quantities in time relative to the dynamics of nuclear reactions involving light elements. In broad brush, these nuclear reactions proceed in the context of a freeze-out from nuclear statistical equilibrium (NSE). In addition, entropy generation from out-of-equilibrium particle decay in such BSM scenarios will dilute the neutrino radiation density and decrease the relativistic degrees of freedom (parameterized by $\neff$). In principle, these alterations from standard cosmology can be calculated with a self-consistent treatment of neutrino energy spectra, plasma temperature, and the strong, electromagnetic, and weak nuclear reactions. A comparison of $\neff$ and light element abundances (principally deuterium and helium) emerging from this epoch with observationally-inferred values of these allows us to explore BSM physics and in some cases to make constraints on the model parameters of this new physics. This program is all the more alluring given anticipated high precision Stage-4 cosmic microwave background (CMB) experiments~\cite{Abazajian:2016yjj} ($\neff$ and primordial ${^4{\rm He}}$) and the advent of 30-m class telescopes~\cite{Maiolino:2013bsa, Cooke:2016rky} (deuterium, hereafter D). The possibility of increased precision in these measurements holds out the promise of better probes of BSM and dark sector physics.

While there are several portals to dark sector physics, we choose to concentrate here on the kinetic mixing portal of a dark photon~\cite{galison1984two, Holdom:1985ag}. This portal is tractable and relatively simple, yet possesses a potentially rich phenomenology of outcomes. Exploration of this portal has gained popularity in recent years. In part, this is because a dark photon could be a dark mediator between a dark sector and SM particles~\cite{Boehm:2003hm, ArkaniHamed:2008qn}. This physics may also explain the muon $g-2$ anomaly~\cite{Pospelov:2008zw}. Moreover, standard model photons manifest as collective modes (e.g., plasmons) in medium. These collective plasma effects can enable resonant (e.g., enhanced) dark photon-photon inter-conversion. In turn, this could produce unique and fruitful dark sector signatures in various plasma environments ranging from compact objects to the early universe~\cite{Redondo:2008aa, Redondo:2008ec, Jaeckel:2008fi, Mirizzi:2009iz, Redondo:2013lna, An:2013yfc, Kunze:2015noa, Hardy:2016kme, Pospelov:2018kdh, Rrapaj:2019eam, McDermott:2019lch, Witte:2020rvb, Caputo:2020bdy, Caputo:2020rnx, Caputo:2020avy}.

For our purposes the dark photon will be a new $U(1)^\prime$ vector particle that has a kinetic mixing with the SM photon~\cite{galison1984two, Holdom:1985ag}. Below the electroweak energy scale the relevant low-energy vacuum Lagrangian we adopt for the dark and standard model electromagnetic sectors is
\beq
	\mathcal{L} \supset - \frac{1}{4} F_{\mu\nu} F^{\mu\nu}- \frac{1}{4} F^\prime_{\mu\nu} F^{\prime\mu\nu} + \frac{\kappa}{2}F_{\mu\nu}F^{\prime\mu\nu} 
	+ \frac{1}{2}m_{A^\prime}{}^2 A^\prime_\mu{}A^{\prime \mu},
	\label{eq: vacuum lagrangian}
\eeq
where here, and hereafter, vector potential $A_\mu$ and field tensor $F^{\mu\nu}$ will refer to the SM photon and electromagnetic fields, while the primed versions, $A^\prime_\mu$ and $F^{\prime\mu\nu}$, will refer to the corresponding dark photon field. Here $\kappa$ is the vacuum kinetic mixing parameter and $m_{A^\prime}$ is the dark photon mass. As for the origin of $m_{A^\prime}$, it could come from a new Higgs mechanism with new light degrees of freedom, or from the Stueckelberg mechanism in which a very heavy dark sector Higgs boson has been integrated out from the theory. In this work, we will consider only the Stueckelberg mechanism alternative.

Previous work investigating the effect of dark photon decay on light element synthesis in the early universe suggested constraints on ranges of dark photon mass and coupling parameter space~\cite{Fradette:2014sza, Berger:2016vxi}. Specifically, the authors in Ref.~\cite{Fradette:2014sza} consider the electromagnetic and hadronic energy injection from dark photon decay, the subsequent photo-dissociation of light nuclei, and the creation of a neutron excess. Their abundance-derived bounds are based on the following features of their calculations: (1) ${\rm D}$ and ${}^4{\rm He}$ are under-produced relative to standard model cosmology, a consequence of out-of-NSE (\lq\lq post-BBN\rq\rq) destruction of nuclei when the dark photon decays to $e^+ e^-$; (2) ${\rm D}$ and ${}^4{\rm He}$ over-production stemming from an increase in the $n/p$-ratio when the dark photon decays to $\pi^+\pi^-$ or $K^+K^-$ prior to NSE-freezeout, $T > 100\,{\rm keV}$; and (3) ${\rm D}$ over-production from an injection of neutrons facilitated by dark photon decay. Comparing the calculated yields of ${\rm D}$, ${}^{3}{\rm He}$, and ${}^{4}{\rm He}$ with the precision measurements of the ${\rm D/H}$ ratio from the high-redshift quasar absorption systems~\cite{pettini2012new, Cooke:2013cba}, they exclude several regions in the model parameter space as shown in figure~\ref{fig:bound}.

A more detailed calculation for the fraction of photons induced from dark photon decay capable of photo-dissociating ${\rm D}$ and ${}^4{\rm He}$ has been shown in Ref.~\cite{Berger:2016vxi}. In that paper, the authors show that the amounts of photo-dissociated ${\rm D}$ and ${}^4{\rm He}$ nuclei stemming from dark photon decay are lower than the amounts reported in Ref.~\cite{Fradette:2014sza}. They conclude that demanding that the ${}^3{\rm He}/{\rm D}$ ratio not exceed the observational limits does not lead to a constraint on the dark photon model parameter space.

In addition to the dark photon considerations in Refs.~\cite{Fradette:2014sza, Berger:2016vxi}, a generic scenario of sub-GeV particles and electromagnetic energy injection from their decays has also been studied, for example in Refs.~\cite{Kawasaki:2020qxm, Forestell:2018txr, Coffey:2020oir, Hufnagel:2018bjp}. In particular, Ref.~\cite{Forestell:2018txr} has included a full electromagnetic cascade (photon and electron) to study the photo-dissociation of the primordial light elements abundances in the post-BBN epoch ($T<10\,{\rm keV}$). This study suggests constraints on late-decaying particles with lifetimes larger than $10^4\,{\rm sec}$ (i.e., post-BBN epochs). Expanding on Ref.~\cite{Forestell:2018txr}, the authors of Ref.~\cite{Coffey:2020oir} have updated their discussions on electromagnetic injection by taking into account non-universality in the photon cascade spectrum relevant for BBN. Reference~\cite{Hufnagel:2018bjp} similarly incorporates an electromagnetic cascade, but additionally evaluates the effects of a non-standard Hubble rate and baryon-to-photon ratio on the predicted primordial nuclear abundances.

In this work, we follow the general scheme of Refs.~\cite{Fradette:2014sza, Berger:2016vxi, Forestell:2018txr, Coffey:2020oir, Hufnagel:2018bjp}, albeit with a completely self-consistent treatment of nuclear reactions and temperature-time-scale factor phasing and entropy flow between the neutrinos and the photon-electron/positron-baryon plasma, as described above. However, we do not treat the post-BBN epoch ($T < 10\,{\rm keV}$), and we consider dark photon masses only between $2\,{\rm MeV}$ and $200\,{\rm MeV}$. In that range of dark photon rest masses, the dominant decay product is into $e^-/e^+$-pairs. We are interested in the range of dark photon kinetic mixing in which the dark photons are produced abundantly early on and, at the same time, have a lifetime such that they decay and inject entropy into the plasma at a time which is in the general time frame of weak decoupling and BBN. In this scenario, the entropy-per-baryon during the BBN epoch will start out with a lower value compared to the CMB-determined value measured at a time (e.g., the recombination epoch at $T \approx 0.2\,{\rm eV}$) well after the end of BBN. Our calculations are then iterated so that out-of-equilibrium dark photon decay injects the right amount of entropy into the plasma to give the correct CMB-determined value. The upshot is that the plasma in the case of dark photon decay starts out ``colder'' than in the case of standard cosmology, with entropy added by decay, altering the time-temperature-scale factor relation over that of standard cosmology. This leads to two consequences, both potentially of use in leveraging comparisons with observations into better probes of dark photon mass and coupling.

The first is a consequence of entropy generated from the decay of dark photons depositing energy into the plasma. This phenomenon has an effect in some ways analogous to $e^+/e^-$ ``annihilation'' in the standard cosmology case. In the standard case, where the entropy in a comoving volume (or entropy per baryon) is constant, the disappearance of the $e^\pm$-pairs in equilibrium means that the entropy they carried is transferred to the photon-electron-baryon plasma, but not to decoupled neutrinos. Hence, eventually the ``temperature'' of the decoupled neutrino component will be lower than that of the photons. Now consider what happens if additionally, because of out-of-equilibrium particle decay, the co-moving entropy is not constant. Entropy generation from dark photon decay dilutes the radiation energy density (as parametrized by $\neff$) and makes the relic neutrinos even colder relative to the photons than in the case of standard cosmology~\cite{Fuller:2011qy}. Some of the calculations we discuss below also include a complete Boltzmann neutrino transport scheme, similarly self-consistently calculated along with all nuclear reactions and dark photon decay processes~\cite{Trans_BBN}. During the protracted weak decoupling and BBN epoch, entropy transfer between the plasma and the decoupling neutrinos is effected by out of equilibrium neutrino scattering on $e^\pm$ pairs, even at temperatures well below $T = 1\,{\rm MeV}$. This late entropy transfer is a small effect in standard cosmology, but may be larger in non-standard scenarios, such as the one we consider here.

The second consequence of out-of-equilibrium dark photon decay arises from the changes in the history of weak, electromagnetic, and strong nuclear reaction rates that accompany entropy injection and dilution. A beginning (lower) value of the entropy-per-baryon during the BBN epoch means that the plasma starts out with a higher value of the baryon-to-photon ratio, $\eta$, compared to the standard cosmology case. As entropy is added by dark photon decay, the time-temperature-scale factor history is altered relative to that in standard cosmology. Entropy is a significant determinant of the abundances of the light elements~\cite{Kolb:1990vq}, affecting both the NSE abundance tracks and the course of the nuclear reactions when the system cools to the point where NSE cannot be maintained and individual nuclear reactions become important. Moreover, the alteration in time-temperature-scale factor also affects the history of the neutron-to-proton ratio, $n/p$. This ratio also influences both NSE and non-equilibrium nuclear reactions. These effects combine to alter light element abundance yields, especially for deuterium and helium, relative to those emerging from a standard cosmology.

The combined effects of dark photon out-of-equilibrium decay may alter each of, for example, $N_{\rm eff}$ and the primordial $^2{\rm H}$ and $^4{\rm He}$ yields in a way characteristic of the dark photon mass and vacuum mixing with the standard model sector. This could enable a comparison of our calculations to high precision observations to provide constraints on, or find signatures of, dark photon physics.

In section~\ref{sec: freeze_in} we discuss the physics of dark photon production and decay in the early universe, specialized for the range of masses and standard model couplings as outlined above. Appendices~\ref{sec: self_energy_function},~\ref{sec: in-medium Lagrangian}, and~\ref{sec: matrix_element}, expand on this physics, with discussions of the polarization tensor, in-medium effects, and dark photon production rates, respectively. In section~\ref{sec: alteration of BBN}, we begin with an exposition of the relevant thermodynamics of the early universe and a discussion of our BBN and neutrino decoupling calculations. We then discuss entropy generation, alterations of neutrino energy spectra, radiation energy density, and nucleosynthesis that accompany dark photon production and decay. Conclusions are given in section~\ref{sec: conclusion}. In what follows we will use the natural units $\hbar=c=1$ throughout this paper unless otherwise specified. The electric charge is $e = \sqrt{4\pi\alpha}\approx 0.303$ with $\alpha \approx 1/137$ being the fine-structure constant.

\section{Thermally-produced dark photons in the early universe}\label{sec: freeze_in}

In this section we describe our calculations for the production and decay of dark photons in the early universe. For the ranges of dark photon masses and couplings to the standard model and the epochs in the early universe we consider here, dark photon equilibrium does not occur. Instead, competition between out-of-equilibrium dark photon production and decay produces an ephemeral freeze-in abundance of these particles. Since the time history of this population of dark photons determines the history of, for example, entropy generation and dilution, and since our constraints are predicated on alteration of observables stemming from those histories, our calculations must accurately capture the physical state of the plasma, neutrino, and dark photon components self-consistently.

\subsection{Pair-annihilation in a dense medium}

Freeze-in dark photons can be produced coherently via photon-dark photon adiabatic conversion or incoherently from interactions with electromagnetic (EM) charge currents. While the former production channel could play an important role in the relic dark photon production at late times, it is suppressed for the case where $m_{A^\prime}>{\rm MeV}$, with peak production occurring before or during BBN. This is because the scattering rate between photon and SM charged particles at the resonance temperature is large enough that the build-up of the relative phase between photon and dark photon states is strongly suppressed~\cite{Redondo:2008ec} -- this is essentially a quantum Zeno effect.

As for the latter production mechanism, there are several possible incoherent production channels, e.g., Compton-like scattering, lepton-pair annihilation into one dark photon, or the same annihilation into one dark photon and one SM photon. When $T\ll m_e$, Compton-like scattering is the dominant channel for dark photon production. When $T\gtrsim m_e$, pair annihilation into one dark photon is the dominant channel. The same annihilation into one dark photon and one SM photon is suppressed by a factor $\alpha \approx 1/137$ due to an additional vertex which makes this channel subdominant. In this work we are interested in the dark photons that are produced abundantly early on, and then decay away during the weak decoupling and BBN epochs. The ranges of dark photon mass and standard model couplings we consider, together with this target range of decay lifetime, picks out epochs with $T\gtrsim m_e$ where lepton- or quark-pair annihilation into one dark photon are the dominant production channels.

In general, the conditions of finite temperature and density characteristic of the early universe plasma will affect the pair-annihilation dark photon production rate~\cite{Braaten:1993jw, Raffelt:1996wa}. The origin of these effects can be put into the language of classical physics: the electric field of the propagating EM wave causes acceleration of free electrons in the medium, altering the dielectric function (dispersion relation) of the EM wave. This in-medium plasma effect produces a standard model (SM) photon self energy. If there is dark photon kinetically mixed with SM photon, then this effect also alters the effective kinetic mixing between SM photon and dark photon. A consequence of this is an enhanced dark photon emission rate.

The plasma medium-induced self-energy of a SM photon is described by adding an additional potential term $-\frac{1}{2}A_\mu\Pi^{\mu\nu}A_\nu$ to the vacuum Lagrangian in equation~(\ref{eq: vacuum lagrangian}), where $\Pi^{\mu\nu}$ is the EM polarization tensor. Rotating away the kinetic mixing term and projecting the vector fields onto one single polarization at a time, we obtain the in-medium Lagrangian for polarization $a$ presented in the mass basis quantities $\tilde{A}_a$ (or $\tilde{F}_{a,\mu\nu}$) and $\tilde{A}^\prime_a$ (or $\tilde{F}^\prime_{a,\mu\nu}$)  as (see appendix~\ref{sec: in-medium Lagrangian})
\beq
\baln
    \mathcal{L}_{{\rm IM},a} \supset &- \frac{1}{4} \tilde{F}_{a,\mu\nu} \tilde{F}^{\mu\nu}_a - \frac{1}{4} \tilde{F}^\prime_{a,\mu\nu} \tilde{F}^{\prime\mu\nu}_a + \frac{1}{2}m_{A'}{}^2 \tilde{A}^\prime_{a,\mu}{}\tilde{A}^{\prime \mu}_a \\
    &+ \frac{1}{2} \pi_a \tilde{A}_{a,\nu} \tilde{A}_a^\nu + e \left(\tilde{A}_{a,\mu} + \frac{\kappa m_{A^\prime}^2}{m_{A^\prime}^2 - \pi_a} \tilde{A}_{a,\mu}^\prime \right) J^\mu_{\rm em},
    \label{eq: in-medium mass basis}
\ealn    
\eeq
where $a$ is either one of the two transverse modes ($\pm {\rm T}$) or the longitudinal mode (L), and $J^\mu_{\rm em}$ is the electric charge current. The function $\pi_a = \pi_a\left(\omega,\mathbf{k}\right)$ is the EM polarization function for the polarization state $a$. Explicit forms for these are given in equations~(\ref{eq: pi_T}) and~(\ref{eq: pi_L}). From the last term in equation~(\ref{eq: in-medium mass basis}), we see that the \emph{effective} coupling between the dark photon $\tilde{A_\mu}^\prime$ and the standard model electric charge current $J_{\rm em}^\mu$ is
\beq
	e\kappa_{{\rm eff},a} = \frac{e\, \kappa\, m_{A'}^2}{\sqrt{ \left(m_{A'}^2 - {\rm Re}\: \pi_a\right)^2 + \left({\rm Im}\: \pi_a\right)^2 } }.
	\label{eq:effective_coupling}
\eeq
Certainly, the physics should be independent of the basis we choose. So in the following discussion, we will refer to the rotated (mass state) $\tilde{A}^{\prime}$ as the dark photon and designate this simply as $A^\prime$.

The physical meaning of the real and imaginary parts of $\pi_a$ follows from considerations of finite temperature field theory. The real part can be interpreted as the effective photon mass in the plasma. With the polarization vectors chosen in equations~(\ref{eq: polarization_T}) and~(\ref{eq: polarization_L}), the dispersion relation for EM waves follows the form $\omega^2 = |\mathbf{k}|^2+{\rm Re}\:\pi_a$ for $a=\pm {\rm T}$ and ${\rm L}$.

The imaginary part of $\pi_a$ describes the rate at which the non-equilibrium dark photon distribution function evolves toward thermal equilibrium. Quantitatively, it is ${\rm Im}\: \pi_a = -\omega \left( \Gamma_{A_a}^{\rm abs} - \Gamma_{A_a}^{\rm prod} \right)$, where $\Gamma_{A_a}^{\rm abs}$ and $\Gamma_{A_a}^{\rm prod}$ denote the absorption rate and spontaneous production rate, respectively~\cite{Weldon:1983jn}. In a local thermal (steady state) equilibrium, detailed balance would dictate that $\Gamma_{A_a}^{\rm prod} = e^{-\omega/T} \Gamma_{A_a}^{\rm abs}$.

\begin{figure}[t!]
\centering
    \includegraphics[width=0.49\textwidth]{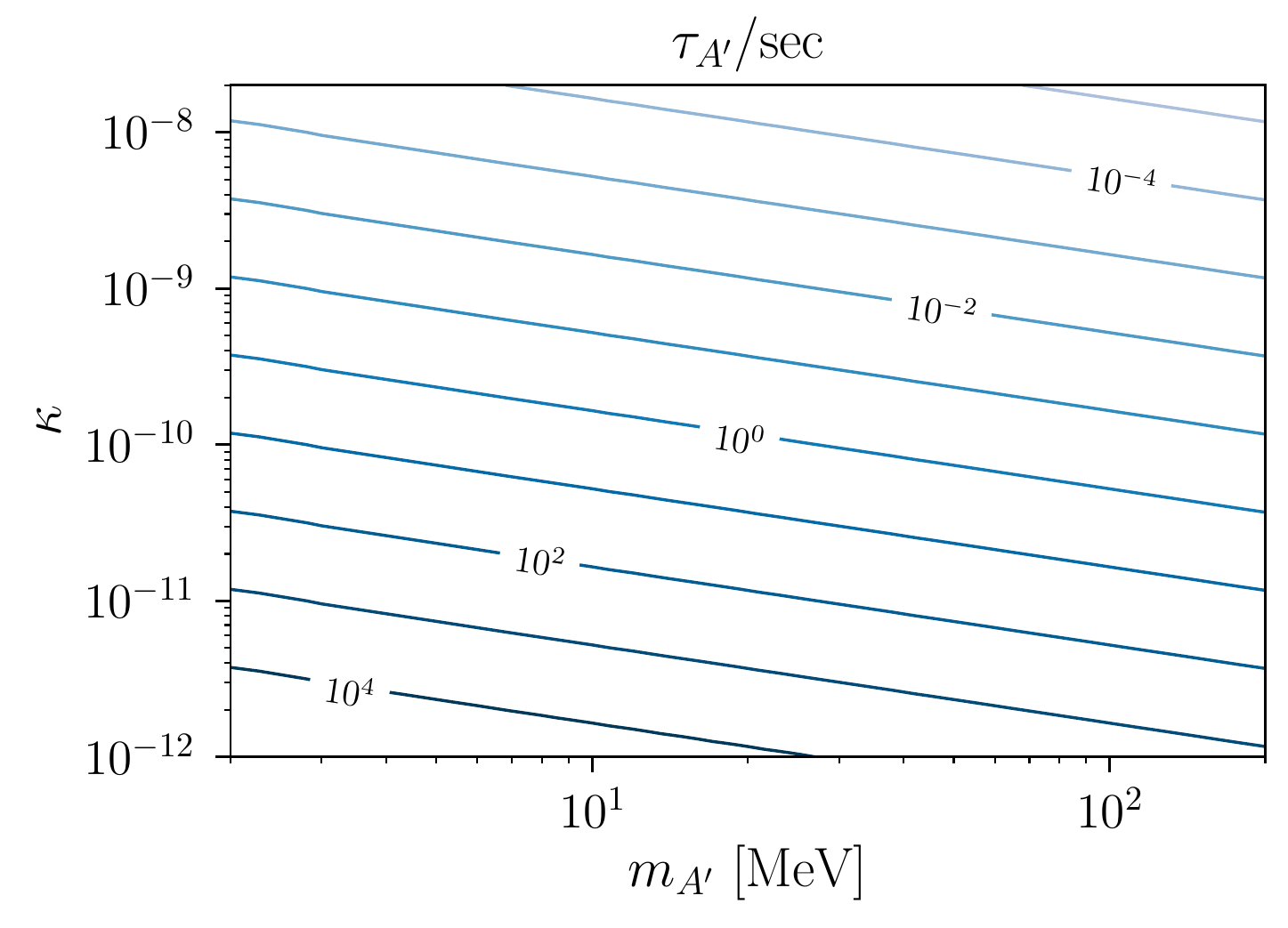}
    \caption{Contours of constant dark photon life time \taudp (in seconds) as functions of $\kappa$ and \massdp.}
    \label{fig:contour_tau}
\end{figure}

Specifically, $\Gamma_{A_a}^{\rm prod}$ in this work denotes the annihilation rate for lepton or quark pairs into one SM photon and is evaluated as (see appendix~\ref{sec: matrix_element})
\beq
\baln
    \Gamma_{A_a}^{\rm prod} \left(\omega\right) = \frac{1}{2\omega} \int &\frac{d^3{\mathbf p}}{(2\pi)^3 \: 2E_{\mathbf p}} \: \frac{d^3{\mathbf q}}{(2\pi)^3 \: 2E_{\mathbf q}} \: \frac{1}{e^{E_\mathbf{p}/T}+1} \: \frac{1}{e^{E_\mathbf{q}/T}+1} \\
    &\sum_\mathrm{spin}  \lvert\mathcal{M}_{l\bar{l} \rightarrow A_a}\rvert^2 \left(2\pi\right)^4 \delta^{(4)} \left(k-p-q\right),
    \label{eq: production rate}
\ealn	
\eeq
where $\mathcal{M}_{l\bar{l} \rightarrow A_a}$ is the matrix element for lepton-pair (momenta ${\bf p}$ and ${\bf q}$) annihilation to one vector boson through a standard EM vertex and the sum is over initial lepton spin states. As a result, the dark photon emission rate in a dense medium is $\kappa_{{\rm eff},a}^2\Gamma_{A_a}^{\rm prod}$. The evolution of the total number density of dark photons can be calculated from the Boltzmann equation as
\beq
	\dot{n}_{A_a^\prime} + 3 H n_{A_a^\prime} = \int \frac{d^3{\mathbf k}}{(2\pi)^3} \: \kappa_{{\rm eff},a}^2 \: \Gamma_{A_a}^{\rm prod}\left(\omega\right) - n_{A^\prime_a} \: \tau_{A^\prime}{}^{-1},
\label{eq:boltzmann}
\eeq
where $H$ is the Hubble parameter and $\tau_{A^\prime}{}^{-1}$ is the dark photon decay rate and is given by~\cite{Fradette:2014sza}
\beq
    \tau_{A^\prime}{}^{-1} = \frac{1}{3} \, \alpha \kappa^2 m_{A^\prime} \left(1+2\frac{m_l^2}{m_{A^\prime}{}^2}\right) \sqrt{1-4\frac{m_l^2}{m_{A^\prime}{}^2}},
    \label{eq:dp_lifetime}
\eeq
with $m_l$ the appropriate lepton rest mass. In figure~\ref{fig:contour_tau}, we show the contours of dark photon lifetime, $\tau_{A^\prime}$, as functions of $\kappa$ and $m_{A^\prime}$.

\subsection{Resonant vs. continuum production}

\begin{figure}[t!]
\centering
    \includegraphics[width=0.49\textwidth]{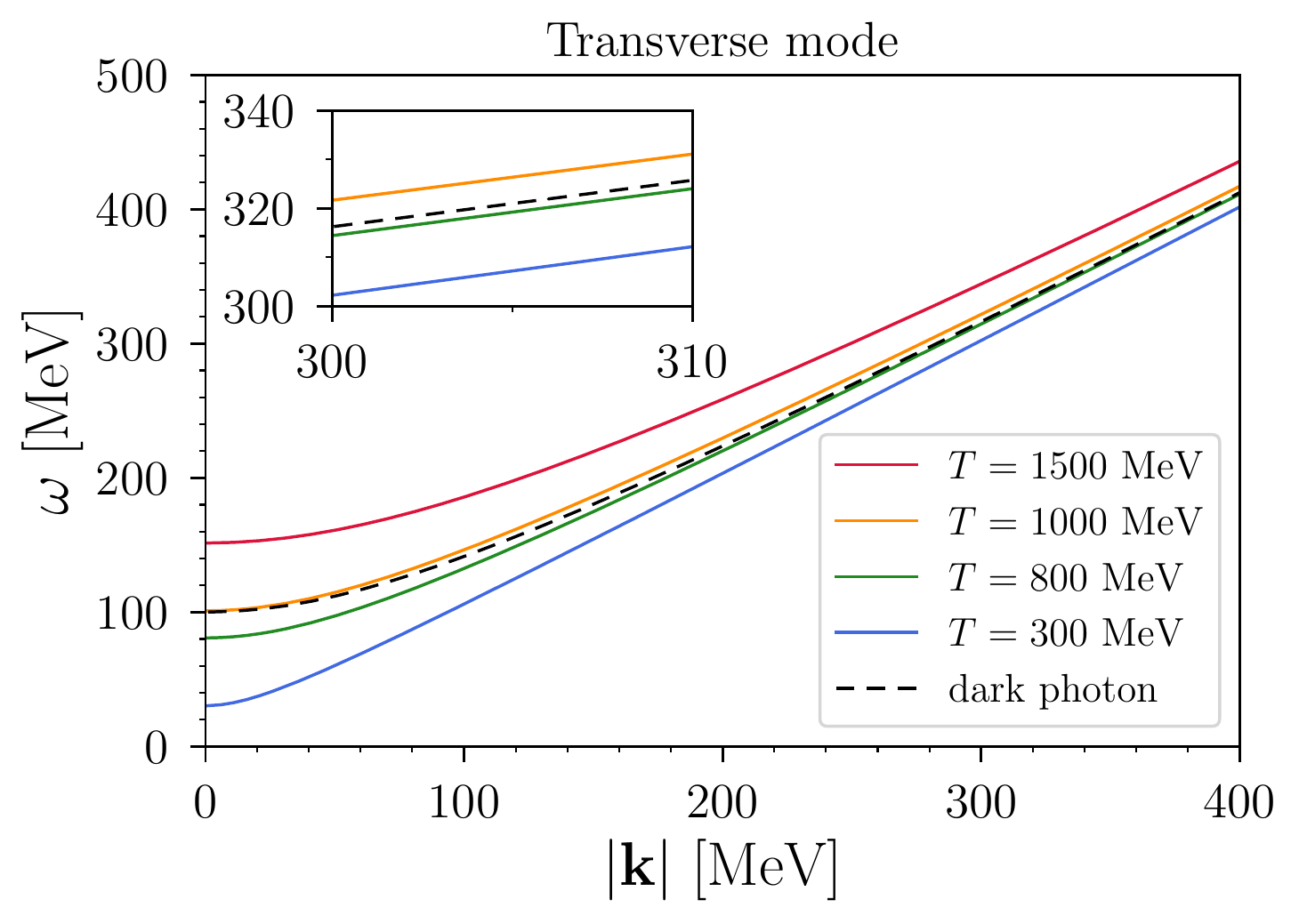}
    \includegraphics[width=0.49\textwidth]{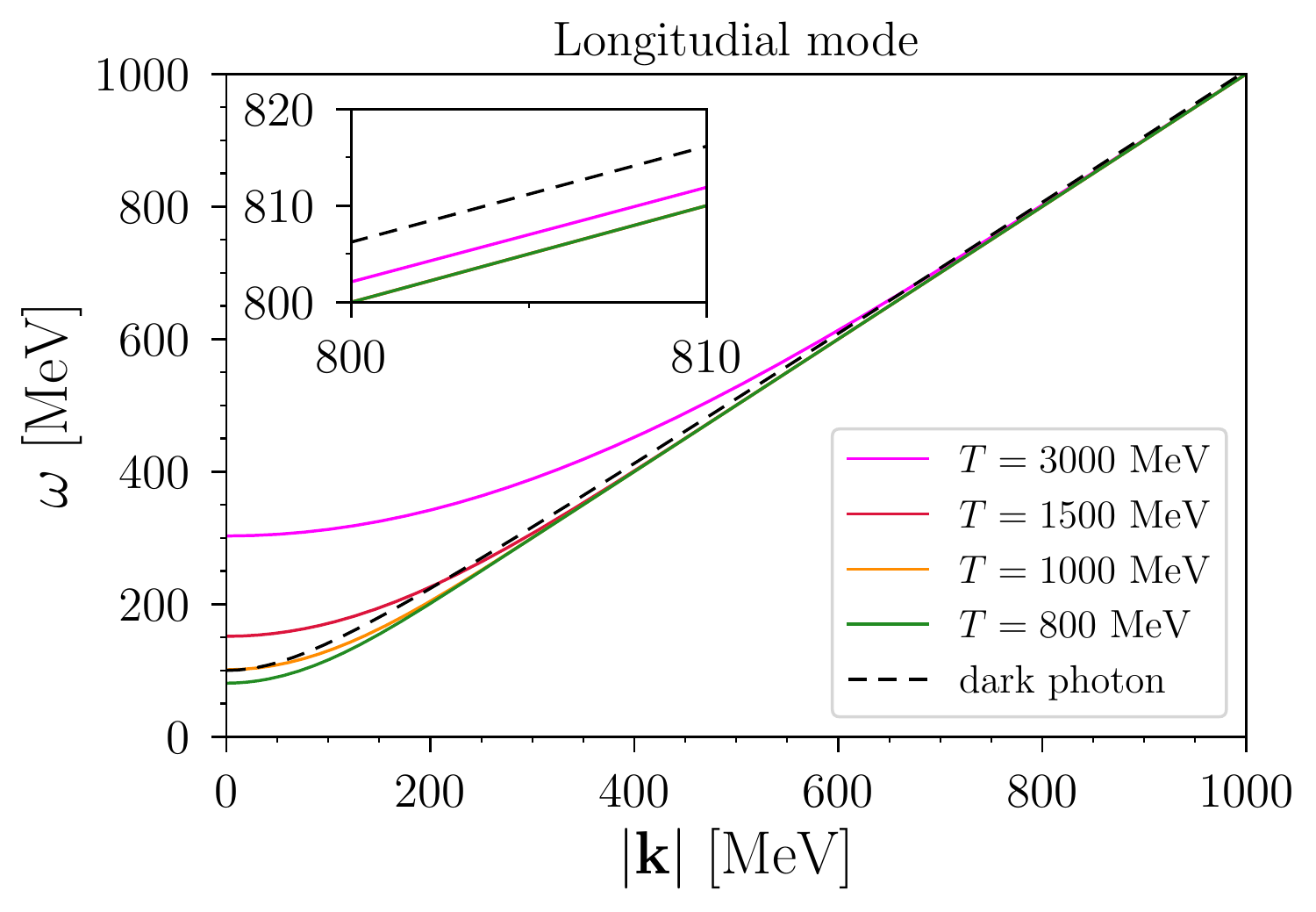}
    \caption{Dispersion relations of the transverse and longitudinal modes for SM photons in a relativistic plasma. The solid curves denote the SM photon dispersion relation at various plasma temperatures. The black dashed curve denotes the dispersion relation of a dark photon model with  $\massdp=100\,{\rm MeV}$. (\textbf{left}) The transverse SM photon dispersion relation curve crosses the dark photon dispersion relation curve in the range  $8\,m_{A^\prime}\lesssim T\lesssim10\,m_{A^\prime}$. Emission of dark photons in either of the two transverse modes is resonantly enhanced in this temperature range. (\textbf{right}) The longitudinal SM photon dispersion relation curve crosses the dark photon dispersion relation curve when $T>10\,m_{A^\prime}$. Emission of dark photons in the longitudinal mode is resonantly enhanced in this range. (\textbf{inset}) The insets show a restricted range in $|\mathbf{k}|$ where the dispersion relations are close to their asymptotic limits with respect to one another.}
    \label{fig: dispersion relation}
\end{figure}

It is clear from the effective coupling expression in equation~(\ref{eq:effective_coupling}) that the dark photon production rate is enhanced when $m_{A'}={\rm Re}\: \pi_a$. Satisfying this \lq\lq resonance\rq\rq\ condition is tantamount to mode matching, requiring a mode solution $(\omega,\mathbf{k})$ such that the dispersion relation for the dark photon is $\omega^2 = |\mathbf{k}|^2 + m_{A^\prime}^2$ and that for in-medium photon mode is $\omega^2 = |\mathbf{k}|^2 + {\rm Re}\, \pi_a\left(\omega,\mathbf{k}\right)$. We can explore the range of temperature conditions in the early universe where the resonance condition can be satisfied by graphically showing the dark photon and in-medium SM photon dispersion relations. We show these in figure~\ref{fig: dispersion relation} for both longitudinal and transverse modes in a relativistic plasma ($T\gg m_e$). From these plots, we see that: (1) resonant dark photon emission of the transverse mode occurs in a narrow range of temperature between $8\,m_{A^\prime}$ and $10\,m_{A^\prime}$; and (2) resonant emission of longitudinal mode occurs at $T \gtrsim 10\,m_{A^\prime}$, which is much higher than the temperature condition for the transverse resonance. In both cases, however, the resonant emission always ceases as $T\lesssim 8\,m_{A^\prime}$.

When the kinetic mixing is off-resonance, i.e., $\lvert m_{A'}{}^2 - {\rm Re}\: \pi_a \rvert \gg  \lvert{\rm Im}\: \pi_a\rvert$, the effective coupling constant becomes $e^2\kappa_{{\rm eff},a}^2 = e^2 {\kappa^2 m_{A'}{}^4} / {\left(m_{A'}{}^2 - {\rm Re}\: \pi_a\right)^2}$. In the low temperature regime where ${\rm Re}\: \pi_a \ll m_{A'}^2 $, the effective kinetic mixing reduces to the vacuum value $\kappa$. The continuum dark photon emission width in this regime is just $\kappa^2 \Gamma_a^{\rm prod}$. In the high temperature regime where ${\rm Re}\: \pi_a \gg m_{A'}^2$, the effective coupling reduces to $\kappa m_{A'}{}^4/{\rm Re}\: \pi_a{}^2$, so the continuum emission rate is suppressed by a factor $m_{A'}{}^4/{\rm Re}\: \pi_a{}^2$ relative to the rate in the low temperature regime~\cite{Hardy:2016kme}. Moreover, there is always more time to produce dark photons at low temperatures than at high temperatures because the Hubble expansion rate in these radiation dominated conditions drops with decreasing temperature, $H\sim T^2/m_{\rm pl}$ with $m_{\rm pl}$ the Planck mass. As a result, the continuum dark photon production is always more significant at low temperatures than at high temperatures.

\begin{figure}[t!]
\centering
    \includegraphics[width=0.5\textwidth]{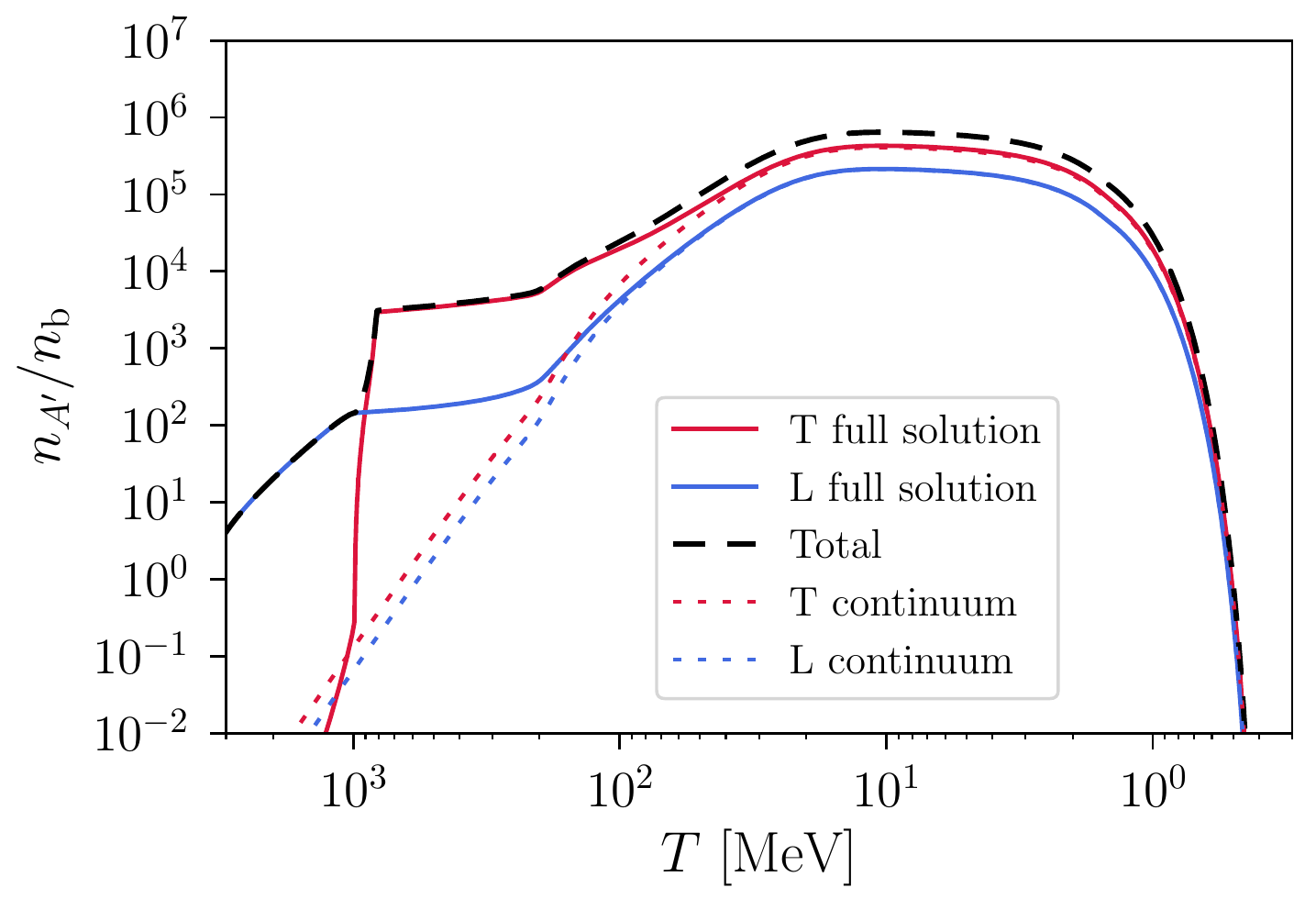}
    \caption{Evolution of the ratio of dark photons to baryons plotted against plasma temperature for a model where $m_{A^\prime} = 100\,{\rm MeV}$ and $\kappa = 10^{-10}$. Red (blue) lines give the transverse (longitudinal) mode. For a given mode, dotted lines show the production history if we ignore plasma effects at all temperatures, i.e., $\kappa_{\rm eff,a}=\kappa$. The result of ignoring the plasma effects gives the continuum contribution. Conversely, solid lines show the complete production history if we include the plasma effects encapsulated in equation~(\ref{eq:effective_coupling}). The dashed black line gives the total number of dark photons for the three modes ($a=\pm {\rm T},{\rm L}$) in the full solution. Resonant production within the plasma occurs at early times ($T\gtrsim 8\, m_{A^\prime}$) while continuum production dominates at late times ($T \lesssim m_{A^\prime}$).}
    \label{fig:production_history}
\end{figure}

We would like to understand the role of the resonant production channel in contributing to the overall dark photon yield, and assess its significance relative to continuum production. As an example, in figure~\ref{fig:production_history} we show the dark photon production history for a specific dark photon mass $m_{A^\prime}=100\,{\rm MeV}$. The solid lines show the full solutions for dark photon emission with in-medium plasma effect included. The solid lines are color coded for longitudinal and transverse modes. On the other hand, the dashed lines show the production histories when no plasma effects are included. The rapid rise in dark photon number density in the temperature range $8\,m_{A^\prime}<T<10\,m_{A^\prime}$, and at $T>10\,m_{A^\prime}$, is a consequence of resonant production of transverse and longitudinal modes. These histories agree with those shown in figure~\ref{fig: dispersion relation}. For the lepton/quark-pair annihilation production channel, we can conclude from the calculations shown in the figure that longitudinal mode resonant production is insignificant relative to resonant transverse mode production. See, for example, Ref.~\cite{An:2013yfc} for a discussion of stellar conditions in the regime where the dark photon mass is less than the plasma frequency, $m_{A^\prime}<\omega_{{\rm p}}$, and where, consequently, the resonant dark photon emission production of longitudinal modes dominates over the resonant transverse mode production rate. On the other hand, the continuum production rates for both transverse and longitudinal modes are initially small at $T > 10\, m_{A^\prime}$ as a consequence of the extra suppression factor $m_{A'}{}^4/{\rm Re}\, \pi_a{}^2$, but these eventually dominate the total dark photon emission when $T \lesssim m_{A'}$. Comparing the full and continuum solutions, we see that: (1) resonant production is important only at $T\gtrsim 8\,m_{A^\prime}$; and (2) eventually the continuum production dominates over the resonant production. Overall, the dark photon yield from the resonant production channels contributes only $\mathcal{O}\left(\lesssim 5\%\right)$ to the total dark photon abundance at $T\approx 0.1\,m_{A^\prime}$. Our calculations employ the same thermal effects on dark photon production as in Ref.~\cite{Fradette:2014sza}, with similar results.

For the numerical simulations presented in the following sections, we include only the continuum emission channels for dark photon production in both the transverse and longitudinal modes. Resonant emission is not included in these calculations.


\section{Alteration of relic neutrino density and nucleosynthesis yield}\label{sec: alteration of BBN}

The key result of out of equilibrium dark photon decay will be to add entropy, altering the time-temperature-scale factor relationship relative to a standard-model-only cosmology. The \emph{final} baryon-to-photon ratio of the universe we live in is a measured quantity. We can infer the entropy per baryon from this quantity. The entropy per baryon, in units of Boltzmann's constant $k_{\rm b}$, for the plasma of electrons, positrons, and photons is 
\beq
\baln
    s_{\rm pl} &= \left(\frac{\pi^4}{45\,\zeta\left(3\right)}\right) \left(\frac{g_{\star S}}{\eta}\right) \\
    &\approx \left(5.91\times 10^9\right) \left(\frac{g_{\star S}}{2}\right) \left(\frac{6.09\times 10^{-10}}{\eta}\right),
    \label{eq:spl_eqn}
\ealn    
\eeq
where $\eta \equiv n_{b}/n_{\gamma}$ is the baryon-to-photon ratio and $g_{\star S}$ is the effective number of degrees of freedom carrying the entropy \cite{Kolb:1990vq}. The PLANCK satellite derives $\eta_{\rm cmb}=6.09\times 10^{-10}$ at the time of recombination ($T\sim 0.2$~eV)~\cite{Aghanim:2018eyx}, which yields $s_{\rm pl,cmb}=5.91\times 10^9$. In standard cosmology with temperature low enough that the baryon number is conserved, $s_{\rm pl}$ is a co-moving invariant. With the presence of entropy injection from dark photon decay, however, the plasma would start out with a lower value of $s_{\rm pl}$ so that its final value at the recombination epoch will match the CMB-determined value, $s_{\rm pl,cmb}$. 

\subsection{Entropy generation and BBN computation}

We use our code \burst \cite{Trans_BBN} to calculate the effects of the production and decay of dark photons during the weak decoupling and BBN epochs. \burst primarily evolves the plasma temperature, neutrino energy spectra, and primordial abundances through these epochs. Adding dark photon physics to this calculation induces three related changes to the standard model case, namely
\begin{enumerate}
  \item A different Hubble expansion rate $H$,
  \item A different plasma temperature versus scale factor (and time) history,
  \item An evolving baryon-density and $s_{\rm pl}$.
\end{enumerate}
To self-consistently follow the three changes we introduce an energy-density variable for the dark photons
\beq\label{eq:rhodp}
  \rhodp = \massdp n_{A^{\prime}},
\eeq
where $n_{A^{\prime}}$ is the total proper number density of dark photons as given by the solutions to equation~(\ref{eq:boltzmann}). In writing equation~(\ref{eq:rhodp}) we have ignored the kinetic contribution to the dark photon energy density. Therefore, our calculations give underestimates of the effects induced by the presence of dark photons. Neglecting the kinetic energy content of the dark photon field makes only small changes, especially where most decays occur for $m_{A^\prime} \gg T_{\rm decay}$. 

We add \rhodp to the energy densities of the other components to calculate the Hubble expansion rate $H$. During dark photon production and decay, we assume the energy density of the electromagnetic plasma instantly equilibrates, which induces a change in the plasma temperature time-derivative \cite{letsgoeu2}
\beq
  \frac{dT}{dt} = -3H\,\frac{\displaystyle\rho_{\rm pl}+P_{\rm pl} +\frac{1}{3H}\frac{dQ}{dt}\biggr|_{T}}{\displaystyle\frac{d\rho_{\rm pl}}{dT}},
  \label{eq:dtempdt}
\eeq
where $\rho_{\rm pl}$ is the energy density of the plasma (less baryons); $P_{\rm pl}$ is the pressure exerted by all plasma components; $dQ/dt|_T$ is the rate of heat gain or lost from nuclear reactions, neutrino scattering/decoupling, and dark photon evolution; and $d\rho_{\rm pl}/dT$ is the temperature derivative of the plasma energy density components (including baryons). We model the energy subtraction (injection) from dark photon production (decay) using the heat sink (source)
\begin{align}
  \frac{dQ}{dt}\biggr|_{T} &= -\frac{dQ}{dt}\biggr|_{\rm nuc}
   + \frac{dQ}{dt}\biggr|_{\nu}
   - \frac{dQ}{dt}\biggr|_{A^{\prime}\leftrightarrow l\overline{l}},\\
  &= -\frac{dQ}{dt}\biggr|_{\rm nuc} + \frac{dQ}{dt}\biggr|_{\nu}
     + \massdp\frac{dn_{A^{\prime}}}{dt}.\\
\end{align}
An injection of heat will raise the entropy per baryon within the plasma $\spl$, which is equivalent to diluting the baryon number density. Therefore, we start with a low entropy-per-baryon and allow the dark photon decays to raise the entropy per baryon (or lower the baryon number density) to a value consistent with photon decoupling, namely $\spl=5.91\times10^9$~\cite{Aghanim:2018eyx}. For each dark photon model (set of dark photon mass and coupling parameters), we iterate on the starting entropy to find the final entropy consistent with Ref.~\cite{Aghanim:2018eyx}, $s_{\rm pl,cmb}$.

\subsection{Neutrino Spectra}\label{sec:nu_spectra}

\begin{figure}[t!]
\centering
    \includegraphics[width=0.5\textwidth]{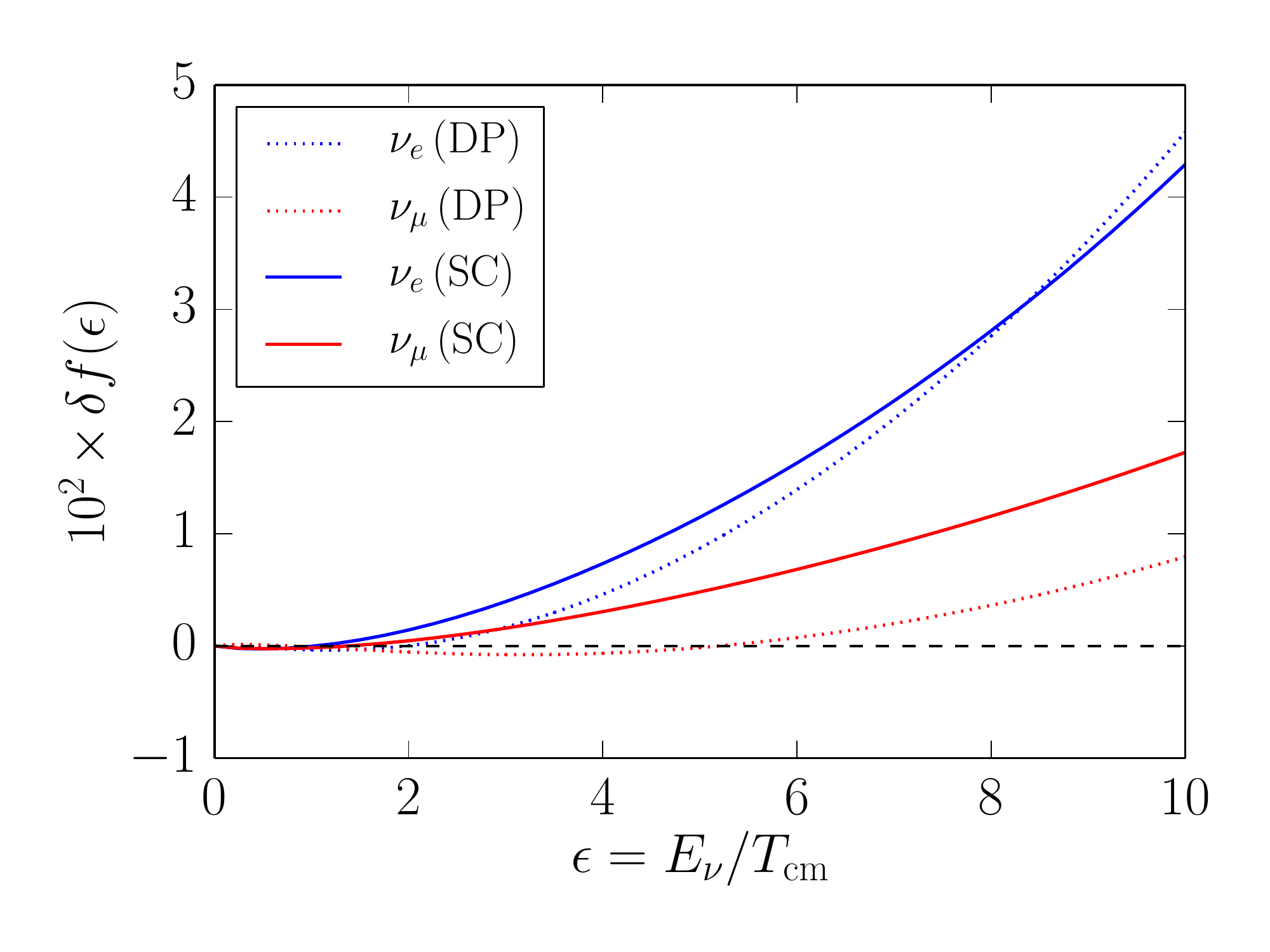}
    \caption{Relative differences from FD [equation~\eqref{eq:rel_fd}] for a dark photon model (dotted) and standard cosmology (solid) versus $\epsilon$ at freeze-out. The parameters for the dark photon model are $\massdp=10\,{\rm MeV}$ and $\kappa=2\times10^{-10}$.}
    \label{fig:spectra}
\end{figure}

As the dark photons decay, they inject heat into the electromagnetic plasma. This heat flow changes the temperature of the plasma giving a different thermal history for the early universe as compared to the standard cosmology. For the dark photon masses we consider in this work, the neutrinos cannot directly partake in this heat flow from dark photon decay. However, a warmer plasma will precipitate a larger heat flow from the plasma into the neutrino seas during  neutrino decoupling. As a result, dark photon decays do affect the neutrino spectra indirectly.

\begin{figure}[t!]
\centering
    \includegraphics[width=0.5\textwidth]{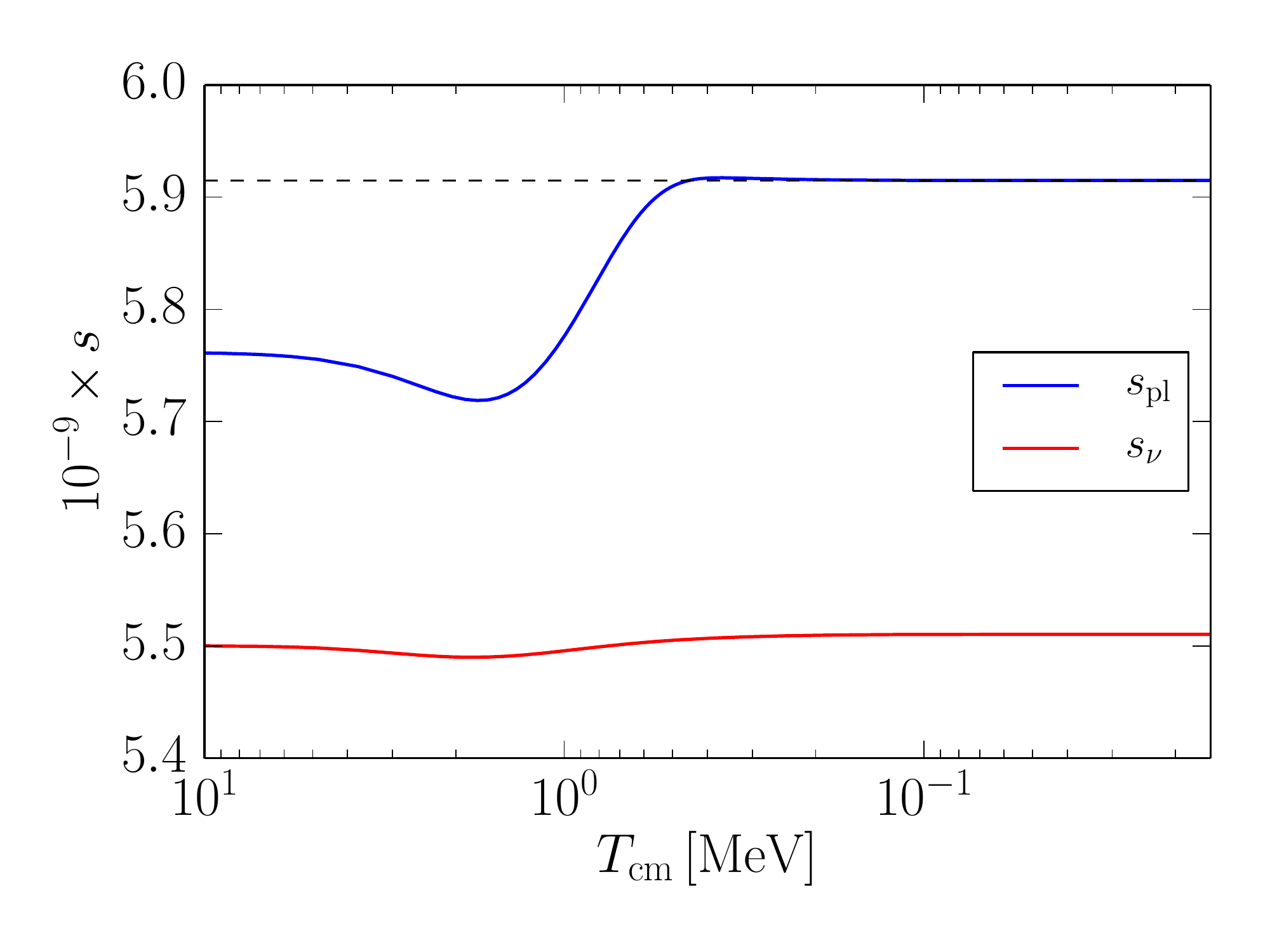}
    \caption{The plasma (blue) and neutrino (red) entropies per baryon versus \tcm for a dark photon model. The parameters for the dark photon model are $\massdp=10\,{\rm MeV}$ and $\kappa=2\times10^{-10}$. The black dashed line is the entropy per baryon as inferred from the CMB in Ref.~\cite{Aghanim:2018eyx}.} 
    \label{fig:sen_v_tcm}
\end{figure}

\begin{figure}[t!]
\centering
    \includegraphics[width=0.5\textwidth]{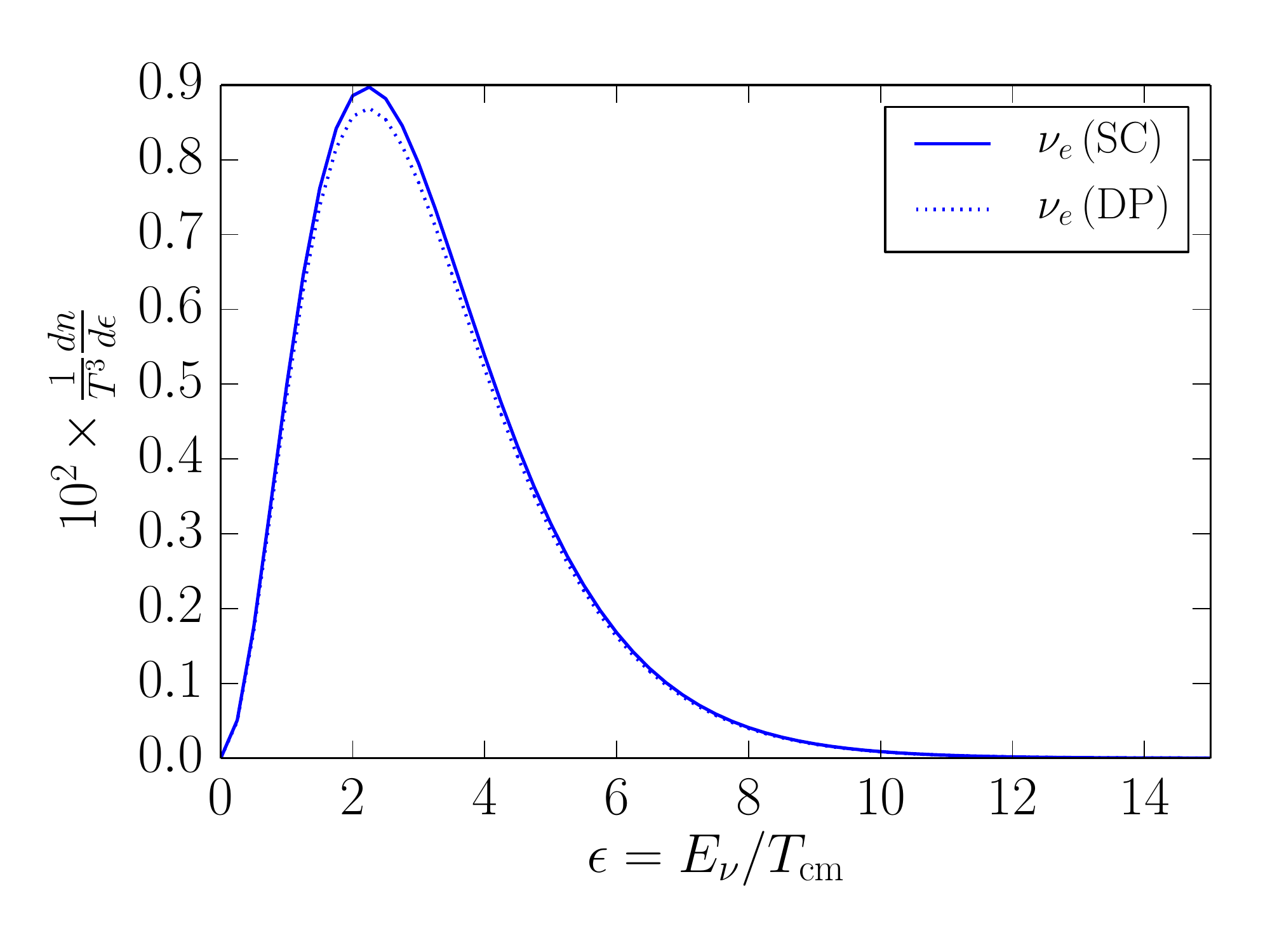}
    \caption{Differential $\nu_e$ number densities scaled by plasma temperature [equation~\eqref{eq:diff_num_dens}] for a dark photon model (dotted) and standard cosmology (solid) versus $\epsilon$ at freeze-out. The parameters for the dark photon model are $\massdp=10\,{\rm MeV}$ and $\kappa=2\times10^{-10}$.}
    \label{fig:dnum}
\end{figure}

As an illustrative example, we take a specific case for dark photon rest mass and coupling to the standard model and calculate in depth how the production and decay of this particle affects weak decoupling and entropy flow. In particular, we show the neutrino energy spectral distortions and the evolution of entropy in figures~\ref{fig:spectra}, \ref{fig:sen_v_tcm}, and \ref{fig:dnum}. For this example case we choose
\beq
    \massdp = 10\,{\rm MeV}, \quad \kappa = 2\times10^{-10},
    \label{eq:dpe}
\eeq
and use the standard cosmological model (i.e., a zero dark photon density) for a baseline comparison. We have picked this particular dark photon model in equation~\eqref{eq:dpe} because of the associated large change in the entropy per baryon during neutrino decoupling. Figure~\ref{fig:spectra} shows the relative changes in the occupation number from FD (Fermi-Dirac) equilibrium
\beq
    \delta f(\epsilon) = \frac{f(\epsilon) - \feq(\epsilon)}{\feq(\epsilon)},
    \quad\feq(\epsilon)=\frac{1}{e^{\epsilon} + 1},
    \label{eq:rel_fd}
\eeq
plotted against the comoving invariant $\epsilon=E_\nu/\tcm$, where $E_\nu$ is the neutrino energy and \tcm is a proxy for (inverse) scale factor \cite{xmelec}. Solid curves give the deviations from FD equilibrium in the case of the standard cosmology, whereas the dotted lines are for the dark photon model in equation~(\ref{eq:dpe}). The blue curves are for the electron-flavor neutrino and the red for $\mu$-flavor. The $\tau$-flavor neutrinos are degenerate with $\mu$-flavor and the antineutrinos are degenerate with the neutrinos in our model of neutrino transport sans oscillations. The black dashed line at zero represents FD equilibrium. The dashed and solid lines deviate from one another, showing two unique histories for neutrino decoupling, one with the dark photon with the assumed parameters, one without.

As the dark photons decay, the entropy increase in the plasma dilutes the neutrino seas and changes the thermal history of the early universe. We show the entropic history for the dark photon decay scenario in figure~\ref{fig:sen_v_tcm}. In this figure, entropy is plotted as a function of the comoving temperature quantity, $T_{\rm cm}$. The blue curve gives the entropy per baryon in the plasma, \spl, and the red curve the entropy per baryon residing in the neutrino seas, \snu. We calculate the plasma entropy from equilibrium thermodynamics. The neutrino seas are out-of-equilibrium so we calculate that entropy using non-equilibrium statistical mechanics, i.e., Boltzmann neutrino energy transport (see section~IV in Ref.~\cite{Trans_BBN}). Both quantities count the number of microstates available to the two subsystems. The dashed black horizontal line in figure~\ref{fig:sen_v_tcm} is the entropy-per-baryon inferred from Ref.~\cite{Aghanim:2018eyx}. There is a small increase in \snu arising from neutrino transport and equivalently encapsulated in the dotted curves of figure~\ref{fig:spectra} at freeze-out. This small increase is accompanied by a small decrease in \spl which is dwarfed by the large increase in the entropy from dark photon decay. The phenomenon of dilution is the increase in the ratio of the entropic quantities from early times to late. The change in the entropy gives a nonstandard thermal history for the early universe. We can summarize the thermal history using the ratio of \tcm to $T$ at freeze-out
\begin{align}
  \frac{\tcm}{T}\biggr|_{\rm f.o.} &= 0.7082\quad \massdp=10\,{\rm MeV},\kappa=2\times10^{-10},\\
  \frac{\tcm}{T}\biggr|_{\rm f.o.} &= 0.7138\quad {\rm Standard\ Cosmology\ (SC)}.
\end{align}

Figures~\ref{fig:spectra} and~\ref{fig:sen_v_tcm} show that the neutrinos experience two competing and opposing effects: an increase in the heat flow from the plasma to the neutrino seas at the level of a few percent deviation (figure~\ref{fig:spectra}); and dilution of the neutrino seas at a level of $20\%$ (figure~\ref{fig:sen_v_tcm}). The former effect raises the number of neutrinos at a given energy bin $\epsilon$ and \tcm, which we write as a differential number density
\beq
  \frac{dn_i}{d\epsilon} = \tcm^3\frac{\epsilon^2}{2\pi^2}f_i(\epsilon),
\eeq 
for a given neutrino flavor $i$. The later effect decreases the number of neutrinos with respect to photons which we encode in the ratio of $\tcm/T$. Figure~\ref{fig:dnum} encapsulates both effects, showing a scaled differential number density
\beq\label{eq:diff_num_dens}
  \frac{1}{T^3}\frac{dn}{d\epsilon} = \left(\frac{\tcm}{T}\right)^3
  \frac{\epsilon^2}{2\pi^2}f(\epsilon),
\eeq
plotted against $\epsilon$. We only plot the scaled differential number densities for electron-flavor neutrinos in the dark photon decay scenario (dotted line) and the standard cosmology (solid line). The $\mu$-flavor quantities are qualitatively identical. The scaled differential number density is a scale-dependent quantity, so we plot figure~\ref{fig:dnum} at the respective freeze-out epochs for each scenario which would occur at different $T$ and \tcm.

The previous exposition has delved into the details of neutrino transport with dark photons. For the specific model we considered, the dominant effect on the neutrino number density (and by extension energy density) was dilution. Energy flow from neutrino transport adds on order an $1\%$ increase to the total neutrino energy density. The increase is dependent on the particular model of dark photons. $\mathcal{O}\left(1\%\right)$ contributions may be important in future high-precision modeling of BSM cosmologies and we emphasize the need for such calculation if/when the data warrant it. For the purposes of this work, we will focus on dilution when discussing the dark photon parameter space in its entirety, and discuss sub-dominant transport effects for specific models.

\subsection{Radiation energy density}
The first observable consequence of entropy injection and dilution is decreasing the neutrino radiation energy density (as parameterized by $\neff$) compared to the value predicted in the standard cosmology. In this subsection, we first calculate the dilution effect in the dark photon model and show the changes in $\neff$ for the full model parameter space; this would be for the case without including energy transport between neutrinos and the plasma. We then discuss the effect of neutrino-energy transport on $\neff$ for a few sets of dark photon model parameters and show the non-linear scaling of the $\neff$ correction with either $m_{A^\prime}$ or $\kappa$.

\begin{figure}[t!]
\centering
    \includegraphics[width=0.5\textwidth]{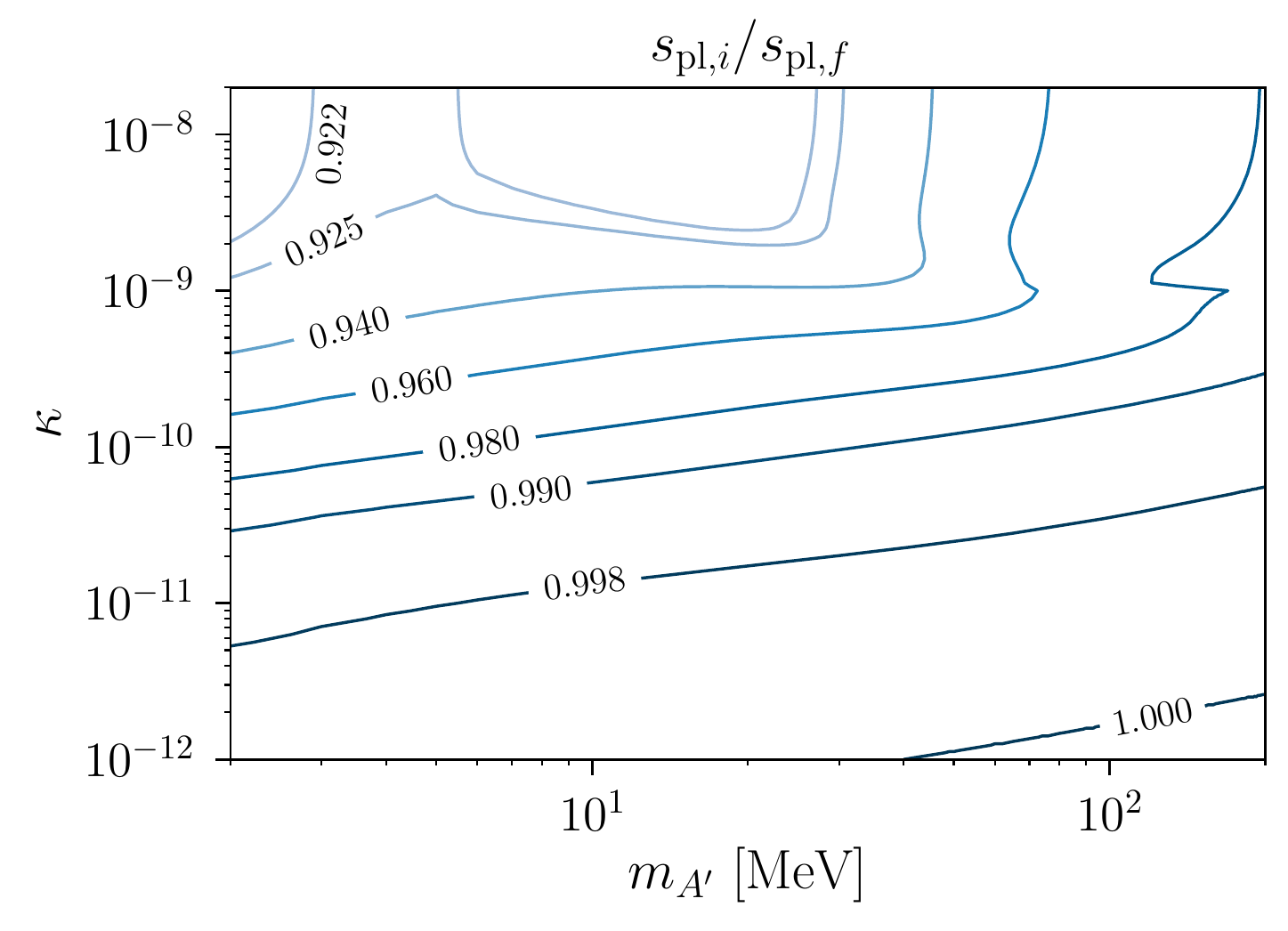}
    \caption{Contours of constant initial-to-final entropy ratios ($s_{{\rm pl},i}/s_{{\rm pl},f}$) plotted in the $\kappa$ versus \massdp parameter space. The contours with values below $1.0$ indicate an increase in entropy due to the dark photon production and decay.}
    \label{fig:contour_sratio}
\end{figure}

\subsubsection{Sharp neutrino decoupling}

The energy density of the neutrino seas is solely a function of \tcm
\beq
  \rho_{\nu} = 6 \left( \frac{7}{8} \right) \left(\frac{\pi^2}{30}\right)\tcm^4
\eeq
when ignoring out-of-equilibrium contributions. The CMB power spectrum is sensitive to the radiation energy density, $\rho_{\rm rad}$, of the early universe, which we parameterize using the quantity \neff and plasma temperature $T$
\beq
  \rho_{\rm rad} = \left[2 + \frac{7}{4}\left(\frac{4}{11}\right)^{4/3}\neff\right]\frac{\pi^2}{30}T^4.
\eeq
If we take the radiation energy density to be the sum of the photon and neutrino components, we find
\beq\label{eq:neff_temps}
  \neff = 3\left(\frac{11}{4}\right)^{4/3}\left(\frac{\tcm}{T}\right)^4.
\eeq

After weak decoupling, dark photon decay injects entropy only into the electromagnetic plasma. This process results in the dilution of both the baryon number and the neutrino energy densities. If $s_{{\rm pl},i}$ is the entropy per baryon in the plasma at an initial epoch, and $s_{{\rm pl},f}$ is the same quantity at a final epoch, then the ratio behaves like the following
\beq
  \frac{s_{{\rm pl},i}}{s_{{\rm pl},f}} = \left( \frac{\frac{2\pi^2}{45}g_{\star S}^{(i)}T_i^3}{\frac{2\pi^2}{45}g_{\star S}^{(f)}T_f^3}\right) \left(\frac{n_{b,f}}{n_{b,i}}\right)= \frac{g_{\star S}^{(i)}}{g_{\star S}^{(f)}}\left(\frac{T_ia_i}{T_fa_f}\right)^3 = \frac{11}{4}\left(\frac{\tcm}{T}\right)^3_{\rm f.o.},\label{eq:ent_ratio}
\eeq
where we have selected the initial epoch such that $T_{{\rm cm},i}=T_i$ and the final epoch such that the ratio $\tcm/T$ has reached a freeze-out value, i.e., all of the plasma entropy resides in SM photons. Figure~\ref{fig:contour_sratio} shows the contours of $s_{{\rm pl},i}/s_{{\rm pl},f}$ in the $\kappa$ vs. \massdp parameter space. All contours are less than or equal to unity, showing that the physics of dark photons precipitates dilution.

\begin{figure}[t!]
\centering
    \includegraphics[width=0.5\textwidth]{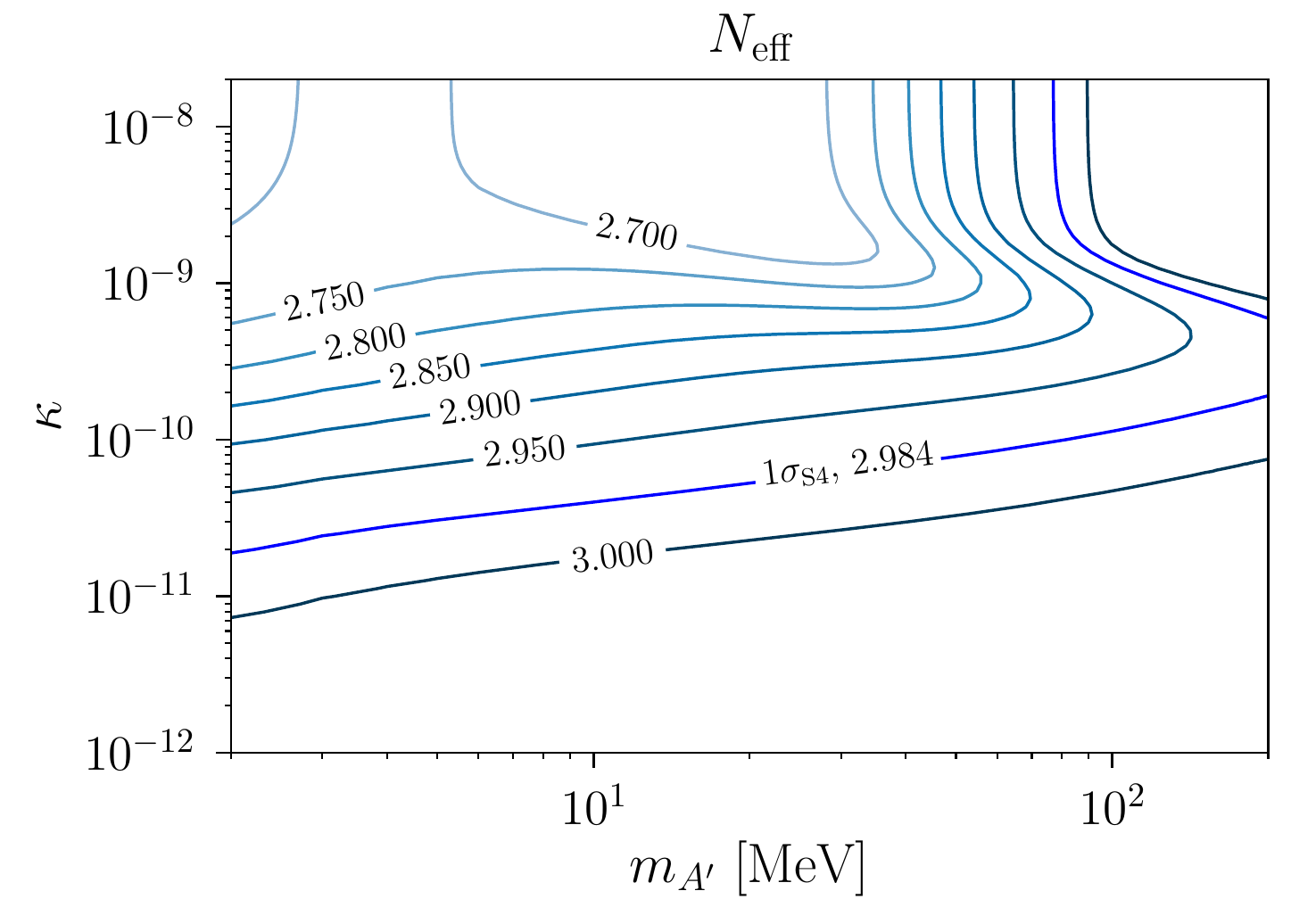}
    \caption{Contours of $N_{\rm eff}$ are shown for values of dark photon mass $m_{A^\prime}$ and mixing parameter $\kappa$. For reference, we also plot the QED-only prediction of $N_{\rm eff}=3.011$ in the absence of neutrino-energy transport. The blue contour is down from $3.011$ by $1\,\sigma_{\rm S4}$ where we quote the measurement uncertainty $1\,\sigma_{\rm S4}=0.027$ from the CMB Stage-4 science book~\cite{Abazajian:2016yjj}.}
    \label{fig:contour_neff}
\end{figure}

If we compare equation~(\ref{eq:ent_ratio}) to equation~(\ref{eq:neff_temps}) evaluated at freeze-out, we find
\beq
  \neff = 3\left(\frac{s_{{\rm pl},i}}{s_{{\rm pl},f}}\right)^{4/3}.
  \label{eq:neff_s}
\eeq
As a result, we expect contours of $N_{\rm eff}$ to correspond directly to the contours of $s_{{\rm pl},i}/s_{{\rm pl},f}$ in figure~\ref{fig:contour_sratio}. That is, a smaller value of $s_{{\rm pl},i}/s_{{\rm pl},f}$ (a larger dilution effect) would lead to a smaller value of $N_{\rm eff}$ (a more diluted neutrino radiation density). Figure~\ref{fig:contour_neff} shows the contours of $N_{\rm eff}$ as functions of $\kappa$ and \massdp in the case of dark photon decay. Indeed, the \neff contours do follow the same general trend of the dilution contours in figure~\ref{fig:contour_sratio}. For low $m_{A^\prime}$, a large value of $\kappa$ induces rapid dark photon production and results in a non-negligible abundance. In addition, peak production occurs in the temperature range $0.1\,m_{A^\prime} \lesssim T_{\rm peak}\lesssim m_{A^\prime}$. For the low end of our mass-range study, peak production occurs after the sharp neutrino decoupling we have instituted for the parameter space scan. This added entropy from dark photon decay dilutes the thermal neutrino seas and lowers $N_{\rm eff}$ to a value smaller than 3. At large $\kappa$ and $m_{A^\prime}\gtrsim 100$~MeV, the dark photons are both created and decay away before neutrino decoupling, and thus there is little or no dilution on the neutrino energy density. The difference in the contour patterns between figures~\ref{fig:contour_sratio} and~\ref{fig:contour_neff} is a result of how we calculate the initial entropy. We fix the initial epoch at $T=30\,{\rm MeV}$ regardless of \massdp. For large \massdp, the entropy is changing in this initial regime and so the respective contours in figure~\ref{fig:contour_sratio} do not meet the criteria used to derive equation~\eqref{eq:neff_s}, and hence diverge from the more precise contours of figure~\ref{fig:contour_neff}.

We plot a blue contour at $\neff=2.984$ on figure~\ref{fig:contour_neff}. This contour uses a $1\,\sigma_{\rm S4}=0.027$ uncertainty in \neff from a CMB Stage-4 forecast \cite{Abazajian:2016yjj}. The $1\,\sigma_{\rm S4}$ difference is between the contour level and the QED-only prediction of $\neff=3.011$ in the absence of heat flow from neutrino-energy transport~\cite{xmelec}. The specific location in the dark-photon parameter space for the $1\,\sigma$ contour would be the same if transport were to add an offset to all of the contour levels, although \neff would take on a value $\approx3.02$ for the $1\,\sigma$ contour in that scenario. However, this procedure relies on the assumption that the effect of transport is independent of the dark photon physics. We expand upon this detail in the following section.

\subsubsection{Effects from neutrino energy transport}

\begin{table*}
  \begin{center}
  \begin{tabular}{c c c c c c}
    \hline
    $m_{A^\prime}$ [MeV] & $\kappa$ & $N_{\rm eff}$ (QED only) & $N_{\rm eff}$ (w/ trans.) & Diff & ${\rm Diff}/\sigma_{\rm S4}$\\
    \midrule[1.0pt]
    SC & & 3.0113 & 3.0442 & 0.0329 & 1.2201\\
    \hline
    2.0 & $1 \times 10^{-12}$ & 3.0097 & 3.0426 & 0.0329 & 1.2192\\
    2.0 & $1 \times 10^{-11}$ & 2.9961 & 3.0289 & 0.0327 & 1.2128\\
    2.0 & $1 \times 10^{-10}$ & 2.8944 & 2.9237 & 0.0293 & 1.0834\\
    2.0 & $1 \times 10^{-9}$ & 2.7201 & 2.7152 & -0.0049 & -0.1838\\
    2.0 & $1 \times 10^{-8}$ & 2.6934 & 2.6838 & -0.0096 & -0.3560\\
    \hline
    10.0 & $2\times10^{-12}$ & 3.0101 & 3.0430 & 0.0329 & 1.2188\\
    10.0 & $2\times10^{-11}$ & 2.9983 & 3.0306 & 0.0323 & 1.1970\\
    10.0 & $2\times10^{-10}$ & 2.9012 & 2.9147 & 0.0135 & 0.5009\\
    10.0 & $2\times10^{-9}$ & 2.7110 &  2.8807 & 0.1697 & 6.2866\\
    10.0 & $2\times10^{-8}$ & 2.6656 & 2.8894 & 0.2238 & 8.2284\\
    \hline
  \end{tabular}
  \end{center}
  \caption{\label{tab:trans}Table of values related to \neff. First and second columns are the dark photon mass and coupling, respectively. Third and fourth columns are the value of \neff with only QED effects and with transport included, respectively. Fifth column is the difference between the fourth and third columns. Sixth column is that difference scaled by the uncertainty in \neff as forecast by CMB Stage-4~\cite{Abazajian:2016yjj}. The first row gives the values calculated in the standard cosmology with our code.}
\end{table*}

The contours of figure~\ref{fig:contour_neff} are for a model of neutrino decoupling which does not include energy transport between neutrinos and charged leptons. In this scenario, the baseline QED-only calculation would yield $\Delta \neff \equiv \neff-3 = 0.011$, where the departure from exactly three is due to finite-temperature QED effects which change the entropy of the plasma~\cite{1994PhRvD..49..611H,1997PhRvD..56.5123F} (see also Ref.~\cite{2020JCAP...03..003B} for a detailed treatment of QED effects in the early universe). In SC calculations of $\neff$ with neutrino transport, the effect of entropy/energy flow from the electromagnetic plasma to the neutrino energy seas increases $\neff$. The sole process of neutrino energy transport yields $0.033<\Delta\neff<0.035$~\cite{Dolgov:1997ne, neff:3.046, Birrell.PhysRevD.89.023008, 2016JCAP...07..051D, 2020JCAP...05..048E, Froustey:2020mcq}. In general, the effects from transport and QED corrections cannot be incoherently summed to give the total change to \neff. The QED effects change the plasma temperature, which changes the rate of heat flow into the neutrino seas, which feeds back on the plasma temperature. However, in practice, this feedback loop is not important at the level of uncertainty in \neff~\cite{Aghanim:2018eyx}, and summing the two contributions gives a range $0.043\lesssim\Delta\neff\lesssim0.046$. A possible interpretation of this result is that the correction on $\neff$ from transport is a constant offset.

We ask the question as to whether transport provides an offset to $\neff$ in the BSM scenarios with dark photons. In Table~\ref{tab:trans}, we show various values of $\neff$ with and without transport for a selection of dark photon masses and couplings. For the models with $\massdp=2.0\,{\rm MeV}$, we see that $\neff$ decreases with increasing $\kappa$ whether transport is included or not. The difference between the two calculations also decreases with increasing $\kappa$, but for $\kappa\ge10^{-9}$, the difference between the two calculations is negative. For these two models, the cooling of the plasma from dark photon production occurs during weak decoupling and induces an entropy flow from the neutrino seas to the plasma -- the reverse of the process in the SC case. When dark photons begin to decay and warm the plasma, weak decoupling has nearly ceased and the neutrino seas do not partake in the increase in radiation energy density. For the models with $\massdp=10.0\,{\rm MeV}$, we see that transport always increases $\neff$ for the range of values of $\kappa$ in Table~\ref{tab:trans}. The difference between the two calculations decreases with increasing $\kappa$, until $\kappa=2\times10^{-10}$, which also is the model studied in detail in section~\ref{sec:nu_spectra}. For models with $\kappa\ge2\times10^{-9}$, transport precipitates larger heat flows than the baseline case of the SC, as shown in the first data row of Table~\ref{tab:trans}. Although we only show full neutrino energy transport calculations for a small region of the parameter space in figure~\ref{fig:contour_neff}, it is clear that corrections from transport scale non-linearly with either \massdp or $\kappa$ and cannot be treated as an offset at the level of future precision. Lastly, we note Ref.~\cite{Ibe:2019gpv} also considers the effects from neutrino energy transport in the dark photon decay scenario. Their $\neff$ result is qualitatively similar to ours.

\subsection{Nucleosynthesis}\label{subsec: Nucleosynthesis}

Another observable consequence of entropy injection and dilution is the alteration of light-element abundance yields. As discussed before, an entropy injection from dark photon decay requires the plasma to start with a lower value of entropy per baryon such that dilution causes $s_{\rm pl}$ to rise to the CMB-determined value, namely $s_{\rm pl,cmb}=5.91\times 10^9$. From the scaling shown in equation~(\ref{eq:spl_eqn}), we see that a lower $s_{\rm pl}$ translates to a higher $\eta$, i.e., the primordial nucleosynthesis environment starts with more baryons in the plasma for the same $T$ than in the case of standard cosmology. This alteration changes the nuclear reaction rates of light-element species relative to the standard cosmology case as the reactions freeze out from the NSE.

\begin{figure}[t!]
\centering
    \includegraphics[width=0.5\textwidth]{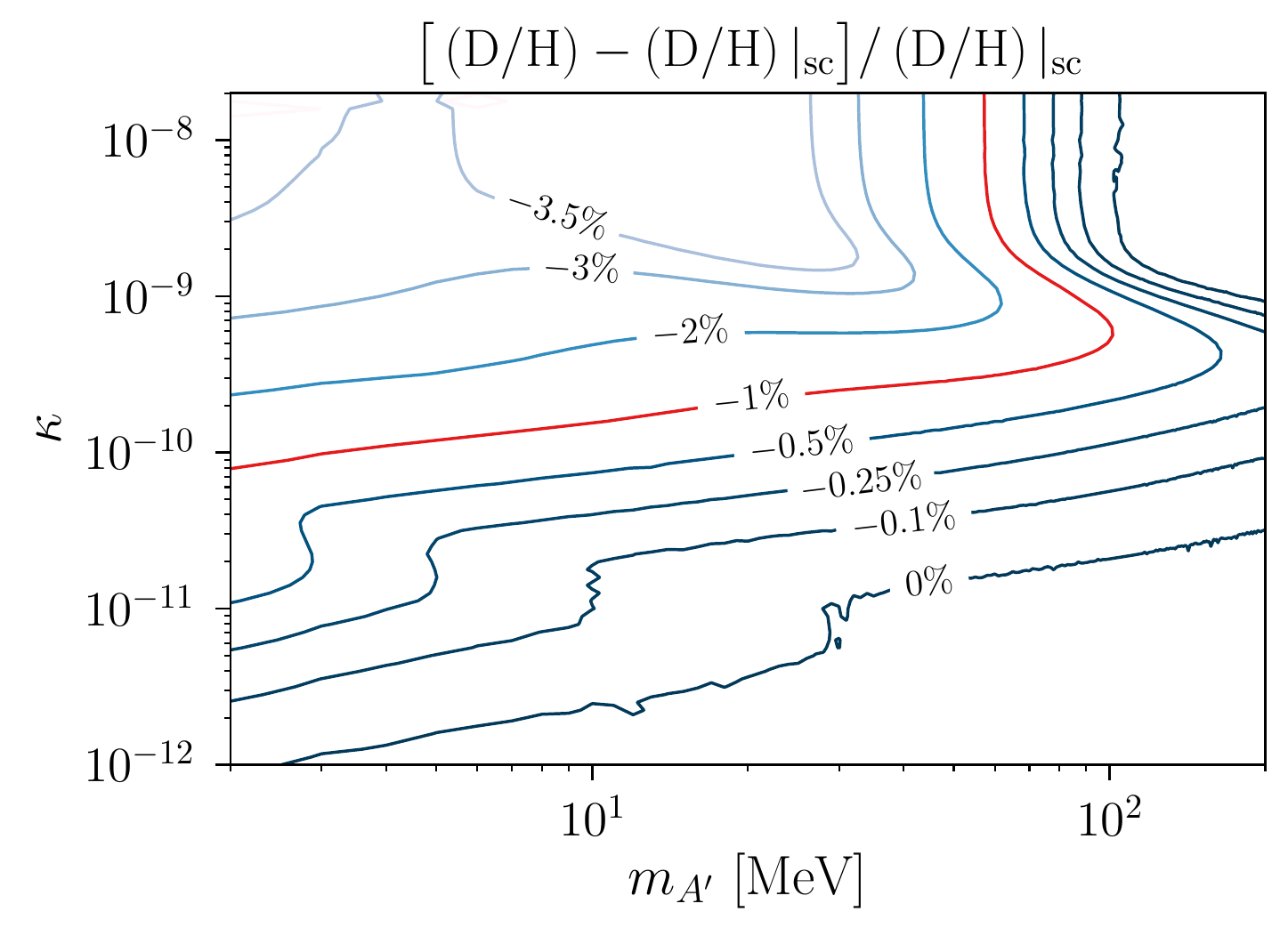}
    \caption{Same parameter space as figure~\ref{fig:contour_neff}, except here we give the percentage change of the primordial deuterium abundance yield in the dark photon model, ${\rm D/H}$, as compared to our calculated standard model physics and standard cosmology result, $\left({\rm D/H}\right)|_{\rm sc} = 2.64 \times 10^{-5}$. The red contour is down from our standard model value by an assumed $1\%$ uncertainty, i.e., $\sigma = 2.64 \times 10^{-7}$. The coarseness of the contour at $0\%$ is a numerical artifact.}
    \label{fig:contour_d}
\end{figure}

At temperatures above $100\,{\rm keV}$, the deuterium abundance is in NSE with the free protons and neutrons. Once the free neutron abundance drops (principally from abrupt alpha particle formation) , deuterium departs from the NSE trajectory. At this point, the evolution of \dtoh proceeds out-of-equilibrium via the nuclear reactions, including but not limited to $n(p,\gamma)d$, $d(p,\gamma)^3{\rm He}$, $d(d,p)t$, and $d(d,n)^3{\rm He}$. A precise determination of the freeze-out (and hence primordial) ratio \dtoh requires data and calculations from ab initio \cite{2015PhRvD..92l3526C,2016PhRvL.116j2501M}, lattice-QCD \cite{2015PhRvL.115m2001B}, experimental \cite{2018PrPNP..98...55B}, and phenomenological \cite{Paris:2014nd,2016ApJ...831..107I,2017ApJ...849..134G,2019PhRvC..99a4619D} sources. The results of those efforts can be integrated into a BBN nuclear reaction network at the appropriate time to yield high-precision absolute BBN predictions. For the dark photon parameter space we study here, we anticipate that changes to \dtoh from updated reaction networks will not depend on the dynamics of dark photon decay, i.e., the effect of an updated network is to linearly perturb a baseline value. As a result, we give our \dtoh results as relative differences from a baseline instead of absolute abundance predictions.

Figure~\ref{fig:contour_d} shows the contours of primordial deuterium abundance yield as functions of mixing parameter and dark photon mass in the case of dark photon decay. The plot is presented as the percentage change of the primordial deuterium abundance in the dark photon model, ${\rm D/H}$, as compared to our calculated standard model and standard cosmology result, $\left({\rm D/H}\right)|_{\rm sc} = 2.64 \times 10^{-5}$. At large $\kappa$ and low $m_{A^\prime}$, dark photons are created abundantly and their decay happens during BBN. That is, the plasma would start out with a lower value of $s_{\rm pl}$ (or higher value of $\eta$) at the BBN epoch than in the case of standard cosmology. This alters the final deuterium abundance yield. At large $\kappa$ and $m_{A^\prime}\gtrsim 100$~MeV, dark photons are both created and decay away too early (well before BBN) to have impact on primordial nucleosynthesis.

We use \dtoh as the diagnostic for BBN in figure \ref{fig:contour_d} because it is well measured and is a priori the most sensitive to changes in entropy. Complementary to \dtoh, the helium mass fraction, \yp, is also well measured and sensitive to the neutron-to-proton ratio $n/p$. The rates of the neutron-to-proton inter-conversion processes dictate the evolution of $n/p$ down to low temperatures. These rates are sensitive to the distributions of neutrinos, anti-neutrinos, electrons, and positrons. In particular, four of these rates are sensitive to the dynamics of dark photons, namely
\begin{align}
  \nu_e + n &\leftrightarrow p + e^-\label{eq:np1}\\
  e^+ + n &\leftrightarrow p + \overline{\nu}_e\label{eq:np2}.
\end{align}
As dark photons begin to decay, the temperature of the plasma increases. The Pauli blocking factors for the charged leptons suppress the forward rate in equation~\eqref{eq:np1} and also the reverse rate in equation~\eqref{eq:np2}. Conversely, the FD occupation factors for the charged leptons enhance the reverse rate in equation~\eqref{eq:np1} and the forward rate in equation~\eqref{eq:np2}. However, the two enhanced charged-lepton capture rates numerically differ from one another because of the mass threshold needed for the reaction to occur, specifically the electron in the reverse reaction in equation~\eqref{eq:np1} must have enough kinetic energy ($\gtrsim0.8\,{\rm MeV}$) to change the proton into a neutron. Increasing the temperature increases the phase space for the electron, implying a larger rate for the reverse reaction in equation~\eqref{eq:np1}. The net effect on $n\leftrightarrow p$ inter-conversion is a slight decrease. We can make a similar argument with respect to the neutrino-capture rates, where the anti-neutrino in the reverse reaction of equation~\eqref{eq:np2} has a threshold of $\sim1.8\,{\rm MeV}$, implying that a suppression in the individual rates leads to a net increase in $n\leftrightarrow p$ inter-conversion. These two effects (along with a change in the Hubble expansion rate) cancel with one another, and $n/p$ in the dark photon scenarios evolves similarly to the SC. There is a net decrease in \yp, but this decrease is at most 1 part in $10^{3}$ and is dwarfed by the change in \dtoh.

Our argument above relies on equilibrium FD distributions for the charged leptons and neutrinos. We showed in Table~\ref{tab:trans} that transport can induce changes in \neff larger than $5\%$. The neutron-to-proton inter-conversion rates are sensitive to the out-of-equilibrium neutrino energy distributions, so the possibility exists that transport can induce larger changes. Indeed, the changes in \yp from transport are an order of magnitude larger than those calculated with equilibrium spectra alone. However, this change is only for the most extreme models in our parameter space, is less than $1\%$, and still remains smaller than the relative changes in \dtoh.

\subsection{Summary of results}

\begin{figure}[t!]
\centering
    \includegraphics[width=0.6\textwidth]{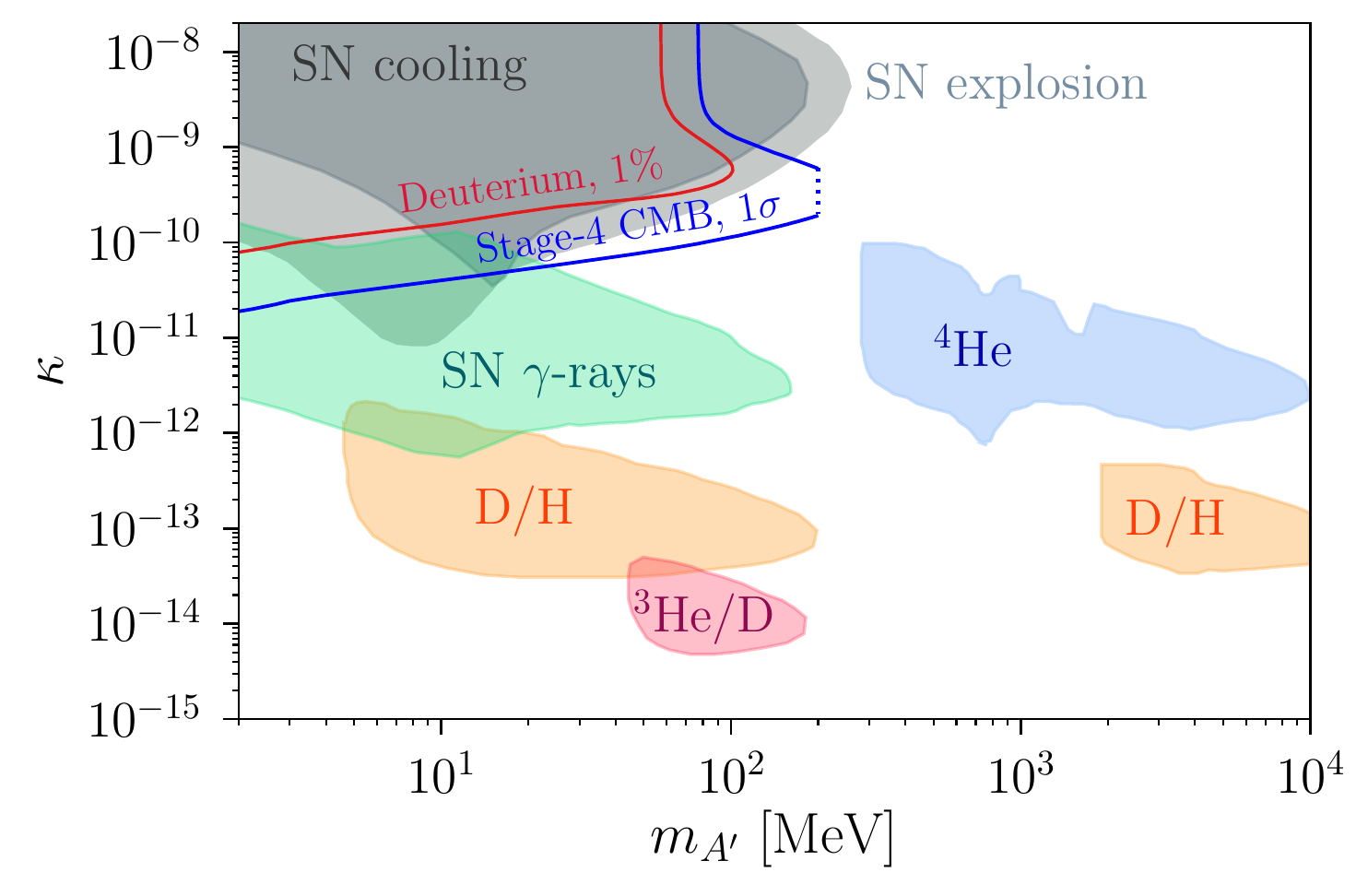}
    \caption{The color-shaded regions show existing bounds on dark photon mass and mixing parameters (as in figures~\ref{fig:contour_neff} and \ref{fig:contour_d}). These bounds, obtained in previous studies, include: SN1987a cooling bound (dark grey)~\cite{Chang:2016ntp}; progenitor envelope bound from core-collapse SN (light grey)~\cite{Sung:2019xie}; non-observation of $\gamma$-rays from SN1987a (green)~\cite{DeRocco:2019njg}; and BBN bounds derived from photo-dissociation and neutron excess (orange, red, blue)~\cite{Fradette:2014sza}. (Note the ${\rm D/H}$ and ${\rm {}^3He/D}$ bounds have been updated in Ref.~\cite{Berger:2016vxi}.) Dark photon parameters lying along the blue line give deviations in $N_{\rm eff}$ which are $1\,\sigma_{\rm S4}$ (where the uncertainty is the CMB Stage-4 science book value, $1\,\sigma_{\rm S4} = 0.027$) below our baseline SC-calculated result with neutrino transport, $\neff=3.044$. Likewise, the red line shows the dark photon parameters giving a $1\%$ deviation of our calculated deuterium yield from our standard model and standard cosmology result, $10^5\times\left({\rm D/H}\right)|_{\rm sc} = 2.64$. The blue dotted line denotes the location  of the upper value of the dark photon mass range, $2\,{\rm MeV}\leq m_{A^\prime}\leq200\,{\rm MeV}$, studied in this work.}
    \label{fig:bound}
\end{figure}

The main results of this work are summarized by the two solid lines in figure~\ref{fig:bound}. The blue line shows the $\neff$ value that is $1\,\sigma_{\rm S4}$ down from our baseline SC calculation of $\neff=3.044$. Here we quote a measurement uncertainty $1\,\sigma_{\rm S4} = 0.027$ from CMB Stage-4~\cite{Abazajian:2016yjj}. Our following conclusion includes the effect of neutrino-energy transport as a constant offset to the dilution physics from dark-photon decay. We caution against applying such an offset to all models. However, Table~\ref{tab:trans} shows that applying an offset is problematic for large values of $\kappa$ which the $1\,\sigma$ contour largely avoids, i.e., the top-left portion of figure~\ref{fig:contour_neff}. We suggest the following interpretation of the blue contour in figure~\ref{fig:bound}. Suppose future CMB Stage-4 experiments measure $\neff$ within $1\,\sigma_{\rm S4}$ down from the SC-predicted value, $\neff=3.044$. From this outcome, we cannot conclude there is entropy generation and dilution from late decay of dark photons during weak decoupling. In that case, we can exclude the narrow range of the model parameter between the blue line and the light grey region (labeled as ``SN explosion'') from the dark photon model since that range of parameter space reduces the $\neff$ by more than $1\,\sigma_{\rm S4}$. As an alternative, suppose CMB Stage-4 experiments measure $\neff$ more than $1\,\sigma_{\rm S4}$ down from $\neff=3.044$. In that case, it suggests there may be late decay of dark photons that cause the dilution of thermal neutrinos during the weak decoupling epoch.

The red line in figure~\ref{fig:bound} shows the deviation of the primordial deuterium abundance yield corresponding to a value of D/H that is down from our calculated standard model physics and standard cosmology value, $\left({\rm D/H}\right)|_{\rm sc}=2.64\times 10^{-5}$, by an assumed $1\%$ uncertainty, i.e., $\sigma=2.64\times 10^{-7}$. This $1\%$-level determination of the primordial deuterium abundance has nearly been obtained via the observation of quasar absorption lines~\cite{Cooke:2017cwo}, which gives $\left({\rm D/H}\right)|_{\rm obs}=\left(2.527\pm 0.030\right)\times 10^{-5}$. The mismatch of the deuterium abundance between our calculated standard model physics result and the measurement reported in Ref.~\cite{Cooke:2017cwo} is due to the uncertainties in our nuclear reaction network which we do not claim as a source of tension. This source of uncertainty restricts us from making constraints on the model parameters with the absolute value of \dtoh. However, we project that uncertainties from the nuclear reaction network will be overcome in the near future (as discussed in section~\ref{subsec: Nucleosynthesis}) and that the ${\rm D/H}$ from the standard model physics and standard cosmology calculation can eventually be directly compared to the observations. Under that assumption, we can make a \emph{potential} bound by applying an assumed $1\%$-level uncertainty to the deuterium abundance yield in our calculation: for the model parameters circled by the red contour line (i.e., the region at the top-left portion of figure~\ref{fig:bound}), the predicted primordial deuterium abundance with the indicated dark photon parameters is under-produced, that is, down by more than $1\%$ from our standard model value. Hence, these dark photon parameters could be potentially ruled out.

We note that while our potential deuterium bound overlaps with the existing supernova bounds~\cite{Chang:2016ntp, Sung:2019xie}, it is obtained from a self-consistent treatment of weak decoupling and BBN physics in the early universe environment. Therefore, our result provides a complementary verification of the supernova bounds derived from the stellar cooling argument.

In figure~\ref{fig:bound}, we also show the existing constraints on the dark photon model as the various color-shaded regions. The dark grey region, labeled as ``SN cooling'', is the bound derived from the anomalous cooling of SN1987a due to the emission of dark photons~\cite{Chang:2016ntp}. The light grey region, labeled as ``SN explosion'', is the bound derived from the energy deposition in the progenitor stellar envelopes via emission of dark photons in the proto-neutron star core~\cite{Sung:2019xie}. The green region shows the bound on the non-detection of gamma-rays, which are produced from the decaying dark photons in supernovae~\cite{DeRocco:2019njg}. These are the dark photon bounds based on an energy argument; they show the constraints on dark photon model parameters for $m_{A^\prime}\lesssim 100\,{\rm MeV}$. Also included in figure~\ref{fig:bound} are the collected BBN bounds from Ref.~\cite{Fradette:2014sza} based on the photo-dissociation of light nuclei and the creation of a neutron excess; they are labeled as ${}^4{\rm He}$, ${}^3{\rm He}/{\rm D}$ and ${\rm D/H}$.

\section{Conclusion}\label{sec: conclusion}

Several conclusions can be drawn from the results of our calculations of dark photon production and decay in the early universe. First, our treatment of these dark photon processes allows for simultaneous, self-consistent calculation of electron-positron-photon-baryon and neutrino physics throughout the weak decoupling and BBN epochs. Dark photons decaying out of equilibrium during this extended period will inject entropy into the medium, leading to time-dependent dilution that modifies how temperature depends on time and expansion, and may modify charged current isospin-changing reactions and the neutron-to-proton ratio history as well. Any such alteration of the time-temperature-scale factor relation from the standard model prediction can result in a concomitant alteration in light element abundance yields and $N_{\rm eff}$. (Note, unlike Refs.~\cite{Fradette:2014sza, Berger:2016vxi}, we do not treat cascade nucleosynthesis but only thermal BBN, and therefore cannot constrain the effects of dark photon decays for temperatures $T<10$~keV.) Exploiting this fact allows us to identify ranges of dark photon mass and couplings with standard model photons that are not currently constrained, but that may be subject to constraint, or probes, with future high precision cosmological data. In particular, CMB Stage-4 measurements promise significant improvements in the precision ($\sim 1\%$) with which the primordial helium abundance and $N_{\rm eff}$ can be measured. Likewise, the primordial deuterium abundance arguably is already known to high precision ($\sim 1\%$), and the advent of 30-m class telescopes promises to increase the confidence in this result.

Key uncertainties in BBN physics remain, for example, in the nuclear reaction rates associated with deuterium production and destruction, and in the effects of quantum kinetic evolution of neutrino flavor during weak decoupling~\cite{2019BAAS...51c.412G}. Calculations of absolute light element abundances and $N_{\rm eff}$ cannot yet attain the projected precision of the observational data. Consequently, we have presented here \emph{deviations} of these quantities relative to our baseline standard model calculations for ranges of dark photon properties. In the end, our results suggest how to extend current limits on these, or even how to look for the ``fingerprints'' of dark photons. The latter would be patterns of deviations in deuterium, helium, and $\neff$ unique to particular ranges of dark photon mass and standard model coupling that are not otherwise constrained.

However, our calculations have revealed an issue which complicates high accuracy calculations of the effects of dark photon production and decay in the early universe. One set of calculations we did, the dark photon parameter survey calculations, are done self-consistently but with an assumption that the neutrino component is completely decoupled. However, neutrinos do not decouple abruptly at the beginning of the weak decoupling epoch, $\sim 10\,{\rm MeV}$. The process of out of equilibrium scattering of neutrinos on $e^\pm$-pairs continues to transfer entropy from the plasma into the decoupling neutrino seas, introducing small distortions in the relic neutrino energy spectra. This is a small effect in standard model cosmology, but can be bigger in non-standard ones. For a set of selected dark photon mass and coupling parameters we have performed fully self-consistent simulations that include full Boltzmann neutrino transport to capture the effects of out of equilibrium neutrino scattering.

As described above, the results of the full transport calculations are sobering. For some ranges of dark photon coupling and mass (the larger values in the ranges we consider) we find that transport can alter the calculated deviations in D/H and $N_{\rm eff}$ from our decoupled-neutrino estimates by of order the projected CMB measurement uncertainties in these quantities. This means that looking for the subtle fingerprints of dark photons (generally lower $N_{\rm eff}$ and D/H but no change in primordial helium) by comparing observational data with calculations will require that those calculations include full scattering-induced neutrino energy transport.

Probing a conjectured dark sector is an alluring prospect, but questions arise. How unique are the fingerprint signatures revealed by the calculations discussed above? Conceivably, a particle other than a dark photon could decay in a way that mimics the entropy injection and dilution history that accompany the production and decay history of a dark photon. Moreover, we have assumed that the dark photon decay is entirely into standard model particles which instantly thermalize. What if the dark sector is rich enough in structure that the dark photon has decay branches into other, presumably lighter, dark sector particles? Were that the case, our calculations would be over-estimates of the entropy injection. Likewise, what if the dark photon, or other dark sector particle, decays during the weak decoupling epoch into standard model neutrinos? Again, entropy injection would be altered but could be tractable with the Boltzmann neutrino transport code described above.

In any case, the weak decoupling and BBN epochs and are promising laboratories for vetting new possibilities for dark sector and BSM physics. Future CMB and deuterium measurements may provide tantalizing clues about a putative dark sector. Here we have shown some of what must be done on the calculation side to translate those clues into insights into new physics.


\begin{acknowledgments}
We thank Susan Gardner and Tongyan Lin for valuable discussions. J. T. L. is supported by a Government Scholarship to Study Abroad from the Taiwan government. We acknowledge the N3AS NSF Hub, supported by NSF Grant No. PHY-1630782 and the Heising-Simons Foundation (2017-228). Additionally, we acknowledge NSF Grant No. PHY-1914242 at UCSD.
\end{acknowledgments}

\appendix 

\section{Electromagnetic polarization tensor}\label{sec: self_energy_function}

In this appendix we review the calculation of the transverse and longitudinal photon polarization functions. Our discussion follows that in Refs.~\cite{Braaten:1993jw, Raffelt:1996wa}. 

For a photon field $A_\mu$ propagating in the $z$ direction with the four-momentum $k^\mu =\left( k^0, \mathbf{k} \right)=\left( \omega, 0 , 0, |\mathbf{k}| \right)$, we choose the basis vectors of the transverse ($\pm{\rm T}$) and longitudinal polarization directions, respectively, as 
\beq
    \hat{e}_{\rm \pm T}^\mu = \frac{1}{\sqrt{2}}\left(0, 1, \pm i, 0 \right),
    \label{eq: polarization_T}
\eeq
\beq
    \hat{e}_{\rm L}^\mu = \frac{|\mathbf{k}|}{\sqrt{\omega^2 - |\mathbf{k}|^2}} \left(1,0,0,\frac{\omega}{|\mathbf{k}|}\right).
    \label{eq: polarization_L}
\eeq    
Each basis vector is normalized, that is, $\hat{e}_a{}^\mu \hat{e}_{a\mu} = -1$ for $a={\rm \pm T}$ or ${\rm L}$. We parametrize the photon polarization tensor in an unmagnetized and isotropic plasma as
\begin{equation}
	\Pi^{\mu\nu}\left(K\right) \equiv \sum_{a=\pm{\rm T},\,{\rm L}} \pi_a\left(\omega, \mathbf{k}\right) \hat{e}_a^{\mu} \hat{e}_a^{*\nu},
\end{equation}
where $\pi_{\rm T}$ and $\pi_{\rm L}$ are transverse and longitudinal polarization functions, respectively. The leading order of $\Pi^{\mu\nu}$ is obtained by evaluating the one-loop photon self-energy insertion and taking the average over the fermion distributions. Approximating the momentum integral by evaluating it at the characteristic fermion velocity, $v_* \equiv \omega_1 / \omega_{\rm p}$, the analytic forms for polarization functions to $\mathcal{O}\left(\alpha\right)$ can be approximated as 
\beq
 	\pi_{\rm T}\left(\omega,\mathbf{k}\right) = \frac{3\omega_{\rm p}^2}{2v_*^2}  \left(  \frac{\omega^2}{|\mathbf{k}|^2} - \frac{\omega^2 - v_*^2 |\mathbf{k}|^2 }{|\mathbf{k}|^2} \frac{\omega}{2v_* |\mathbf{k}|} \ln\frac{\omega + v_* |\mathbf{k}|}{\omega - v_*|\mathbf{k}|}   \right),
 	\label{eq: pi_T}
\eeq 
\beq
	\pi_{\rm L}\left(\omega,\mathbf{k}\right) = \frac{3 \omega_{\rm p}^2 }{v_*^2} \left( \frac{\omega^2 - |\mathbf{k}|^2}{|\mathbf{k}|^2} \right) \left( \frac{\omega}{2 v_*|\mathbf{k}|} \ln\frac{\omega + v_*|\mathbf{k}|}{\omega - v_*|\mathbf{k}|}-1\right),
	\label{eq: pi_L}
\eeq
where 
\beq
	\omega_1^2 \equiv  \frac{4\alpha}{\pi}  \int_0^\infty  dp  \frac{p^2}{E} \left( \frac{5}{3} v^2 -v^4 \right) \big[f_l\left(E\right) + f_{\bar{l}}\left(E\right) \big],
\eeq
\beq
	\omega_{\rm p}^2 \equiv  \frac{4\alpha}{\pi}  \int_0^\infty  dp \frac{p^2}{E} \left( 1 - \frac{1}{3} v^2 \right) \big[f_l\left(E\right) + f_{\bar{l}}\left(E\right) \big].
\eeq
With the conventions used in equations~(\ref{eq: polarization_L}) and (\ref{eq: pi_L}), the dispersion relation is written in the form $\omega^2=|\mathbf{k}|^2+\pi_a\left(\omega,\mathbf{k}\right)$ for $a=\pm{\rm T}$ and ${\rm L}$ modes, and $\omega_{\rm p}$ is the plasma frequency. In these expressions, $f_l$ and $f_{\bar{l}}$ are the lepton and anti-lepton occupation probabilities, respectively.

The electrons and positrons in equilibrium in the high entropy-per-baryon plasma of the early universe plasma are relativistic when $T\gg 1\,{\rm MeV}$ and these particles have negligible chemical potentials (i.e., $T\gg|\mu_{e^+}|,\,|\mu_{e^-}|$). In this limit, all electrons and positrons have velocity $v = v_\star = 1$ and the plasma frequency is $\omega_{\rm p} = \sqrt{4\pi\alpha T^2/9}$.

In the limit of a non-relativistic and neutral electron-proton plasma (e.g., the plasma in the sun), the transverse and longitudinal photon polarization functions are given as
\beq
    \pi_{\rm T}\left(\omega,\mathbf{k}\right) = \omega_{\rm p}^2 \left(1 + v_{{\rm th}, e}^2 \frac{|\mathbf{k}|^2}{\omega^2}\right),
\eeq
\beq
    \pi_{\rm L}\left(\omega,\mathbf{k}\right) = \omega_{\rm p}^2 \left( \frac{\omega^2 - |\mathbf{k}|^2}{|\mathbf{k}|^2} \right) \left(\frac{|\mathbf{k}|^2}{\omega^2} + 3v_{{\rm th},e}^2 \frac{|\mathbf{k}|^4}{\omega^4}\right),
\eeq
where $v_{{\rm th}, e}\equiv \sqrt{T/m_e}$ is related to the electron thermal speed. The plasma frequency in this limit is $\omega_{\rm p} = \sqrt{4\pi\alpha n_e/m_e}$.

We note that another popular convention for the longitudinal basis vector is 
\beq
    \hat{e}_{\rm L}^\mu = \left( 1, 0, 0, \frac{\omega}{|\mathbf{k}|} \right),
\eeq
and its corresponding longitudinal polarization function takes the form
\beq
	\pi_{\rm L}\left(\omega,\mathbf{k}\right) = \frac{3 \omega_{\rm p}^2 }{v_*^2} \left( \frac{\omega}{2 v_*|\mathbf{k}|} \ln\frac{\omega + v_*|\mathbf{k}|}{\omega - v_*|\mathbf{k}|}-1\right).
\eeq
With this convention, the dispersion relation for the longitudinal mode is  $|\mathbf{k}|^2 = \pi_{\rm L}$. However, throughout this paper we follow the convention used in equations~(\ref{eq: polarization_L}) and (\ref{eq: pi_L}).

\section{In-medium effect to dark photon couplings}\label{sec: in-medium Lagrangian}

The effective couplings of a massive dark photon depends strongly on the properties of SM photon polarization in the dense medium. In this appendix we review the in-medium Lagrangian and the conditions required for resonant dark photon emission~\cite{Lin:2019uvt}.

The self energy of the photon field $A^\mu$ in a dense medium is described by including an additional potential term $-\frac{1}{2} A_\mu \Pi^{\mu\nu}A_\nu$ in the vacuum Lagrangian in equation~(\ref{eq: vacuum lagrangian}). After making a field redefinition $A_\mu\to A_\mu+\kappa A_\mu^\prime$ to rotate away the kinetic mixing term, the in-medium Lagrangian of the relevant terms to $\mathcal{O}\left(\kappa\right)$ becomes
\beq
	\mathcal{L}_{\rm IM} \supset - \frac{1}{4} F_{\mu\nu} F^{\mu\nu} - \frac{1}{4} F^\prime_{\mu\nu} F^{\prime\mu\nu}
	+ \frac{1}{2}m_{A'}{}^2 A^\prime_\mu{}A^{\prime \mu} - \frac{1}{2}A_\mu \Pi^{\mu\nu}A_\nu - \kappa A_\mu \Pi^{\mu\nu}A_\nu^\prime + e(A_\mu +\kappa A_\mu^\prime) J^\mu_{\rm em}.
	\label{eq: in-medium Lagrangian}
\eeq
Next, we project the photon and dark photon fields onto transverse ($\pm{\rm T}$) and longitudinal (${\rm L}$) directions and consider only one polarization at a time. This can be done by decomposing a given vector field $V^\mu$ into its three polarization states as 
\beq
    V^\mu = \sum_{a=\pm{\rm T},\,{\rm L}} V_a \hat{e}^\mu_a \equiv \sum_{a=\pm{\rm T},\,{\rm L}} V_a^\mu ,
\eeq
where again each basis vector satisfies $\hat{e}_a{}^\mu \hat{e}_{a,\mu} = -1$. As a result, the in-medium Lagrangian of one given single polarization state $a$ is 
\beq
\baln
    \mathcal{L}_{{\rm IM},a} \supset &- \frac{1}{4} F_{a,\mu\nu} F^{\mu\nu}_a - \frac{1}{4} F^\prime_{a,\mu\nu} F^{\prime\mu\nu}_a + \frac{1}{2}m_{A'}{}^2 A^\prime_{a,\mu}{}A^{\prime \mu}_a  \\
    &+ \frac{1}{2} \pi_a A_{a,\nu}A_a^\nu + \kappa \pi_a A_{a,\mu}A_a^{\prime\mu} + e(A_{a,\mu} +\kappa A_{a,\mu}^\prime) J^\mu_{\rm em}.
\ealn    
\eeq
The mixing between the photon and dark photon fields can be rotated away by making another field redefinition,
\beq
\baln
    A_{a,\mu} &= \tilde{A}_{a,\mu} + \frac{\kappa \pi_a}{m_{A^\prime}^2-\pi_a} \tilde{A}_{a,\mu}^\prime, \\
    A_{a,\mu}^\prime &= \tilde{A}_{a,\mu}^\prime - \frac{\kappa \pi_a}{m_{A^\prime}^2-\pi_a} \tilde{A}_{a,\mu}.
\ealn    
\eeq
Eventually, we arrive at the in-medium Lagrangian of the polarization state $a$ presented in the mass basis as\footnote{We note that the form of effective kinetic mixing presented in equation~(\ref{eq: in-medium Lagrangian mass basis}) works for all three polarization states since they satisfy the same form of normalization, $\hat{e}_a^\mu \hat{e}_{a,\mu} = -1$.}
\beq
\baln
    \mathcal{L}_{{\rm IM},a} \supset &- \frac{1}{4} \tilde{F}_{a,\mu\nu} \tilde{F}^{\mu\nu}_a - \frac{1}{4} \tilde{F}^\prime_{a,\mu\nu} \tilde{F}^{\prime\mu\nu}_a + \frac{1}{2}m_{A'}{}^2 \tilde{A}^\prime_{a,\mu}{}\tilde{A}^{\prime \mu}_a \\
    &+ \frac{1}{2} \pi_a \tilde{A}_{a,\nu} \tilde{A}_a^\nu + e \left(\tilde{A}_{a,\mu} + \frac{\kappa m_{A^\prime}^2}{m_{A^\prime}^2 - \pi_a} \tilde{A}_{a,\mu}^\prime \right) J^\mu_{\rm em}.
    \label{eq: in-medium Lagrangian mass basis}
\ealn    
\eeq
It is clear from equation~(\ref{eq: in-medium Lagrangian mass basis}) that the effective coupling between $\tilde{A}^\prime_a$ and $J_{\rm em}$ is
\beq
    e\kappa_{{\rm eff},a} = \frac{e\kappa m_{A'}{}^2}{\sqrt{ \left(m_{A'}{}^2 - {\rm Re}\: \pi_a\right)^2 + \left({\rm Im}\: \pi_a\right)^2 } },
\eeq
and the dark photon emission rate will be enhanced when ${\rm Re}\:\pi_a$ approaches $m_{A'}{}^2$.

\subsection{Example: resonant dark photon emission in a nonrelativistic plasma}
References~\cite{An:2013yfc, Redondo:2013lna, Redondo:2008aa} have pointed out the importance of plasma effects in the dark photon emission rate in the sun and in horizontal branch stars when $m_{A^\prime}<10$\,eV. Here we use the plasma dispersion relation to interpret these results.

In compact objects, the electron plasma frequency is many orders of magnitude higher than electron cyclotron frequency. As far as the ordinary electromagnetic (transverse) and electrostatic (longitudinal) modes are concerned, the plasma in such conditions can be treated as unmagnetized and isotropic. A SM photon propagating in this environment would then acquire an effective in-medium mass, ${\rm Re}\,\pi_a$, where the general form of $\pi_a$ is given in equations~(\ref{eq: pi_T}) and (\ref{eq: pi_L}). With the presence of a dark photon with mass $m_{A'}$, dark photon resonant emission occurs when $m_{A'}{}^2 = {\rm Re}\,\pi_a$. This statement is equivalent to saying that the resonance happens when there is a solution of $(\omega,\mathbf{k})$ that satisfies both the dispersion relations for the dark photon, $\omega^2 = |\mathbf{k}|^2 + m_{A^\prime}^2$, and for in-medium SM photons, $\omega^2 = |\mathbf{k}|^2 + {\rm Re}\, \pi_a\left(\omega,\mathbf{k}\right)$. While these two dispersion relations are similar in structure, they dictate quite different behavior in a nonrelativistic plasma such as that in the sun or in horizontal branch stars.

\begin{figure}[t!]
\centering
    \includegraphics[width=0.49\textwidth]{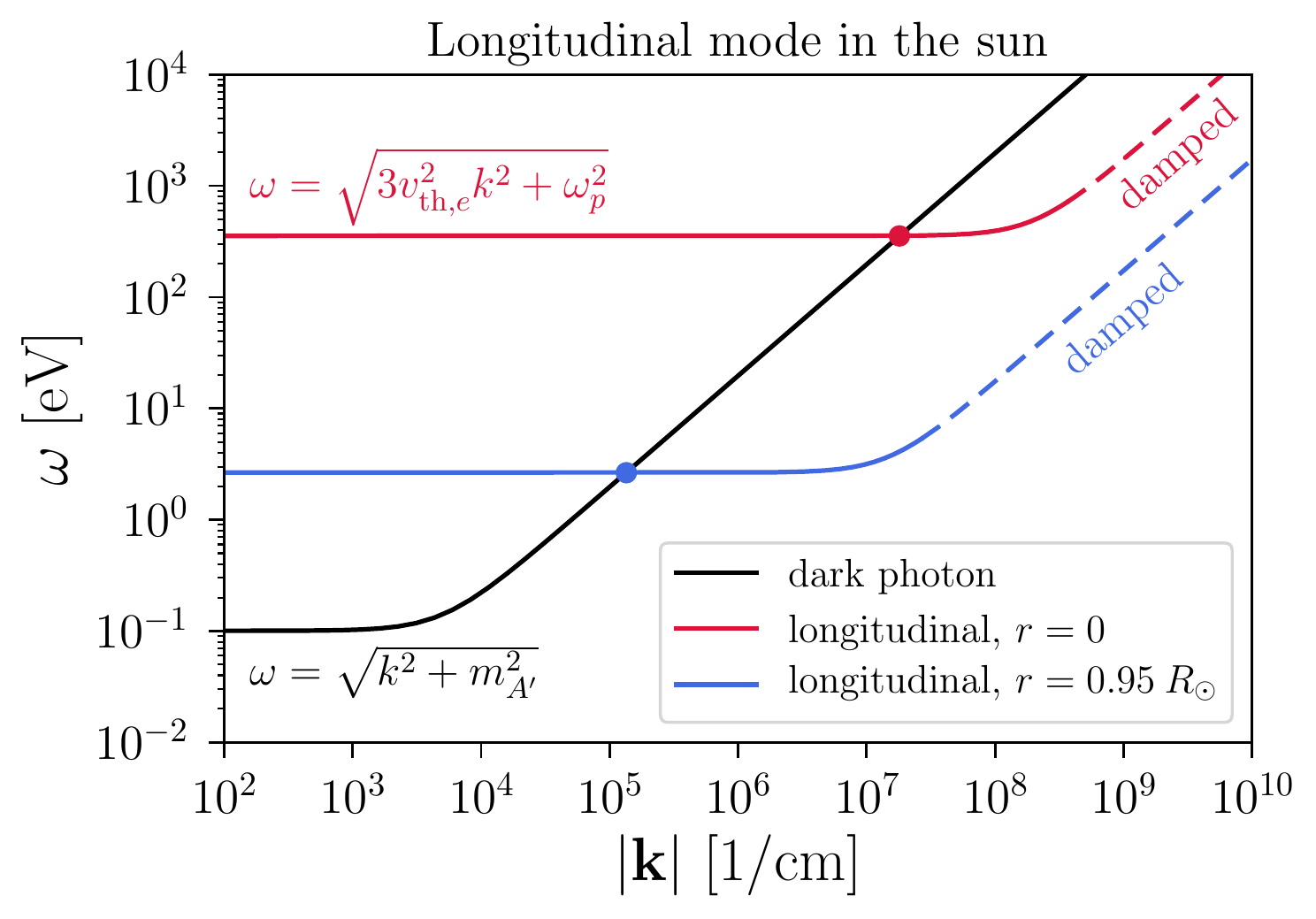}
    \includegraphics[width=0.49\textwidth]{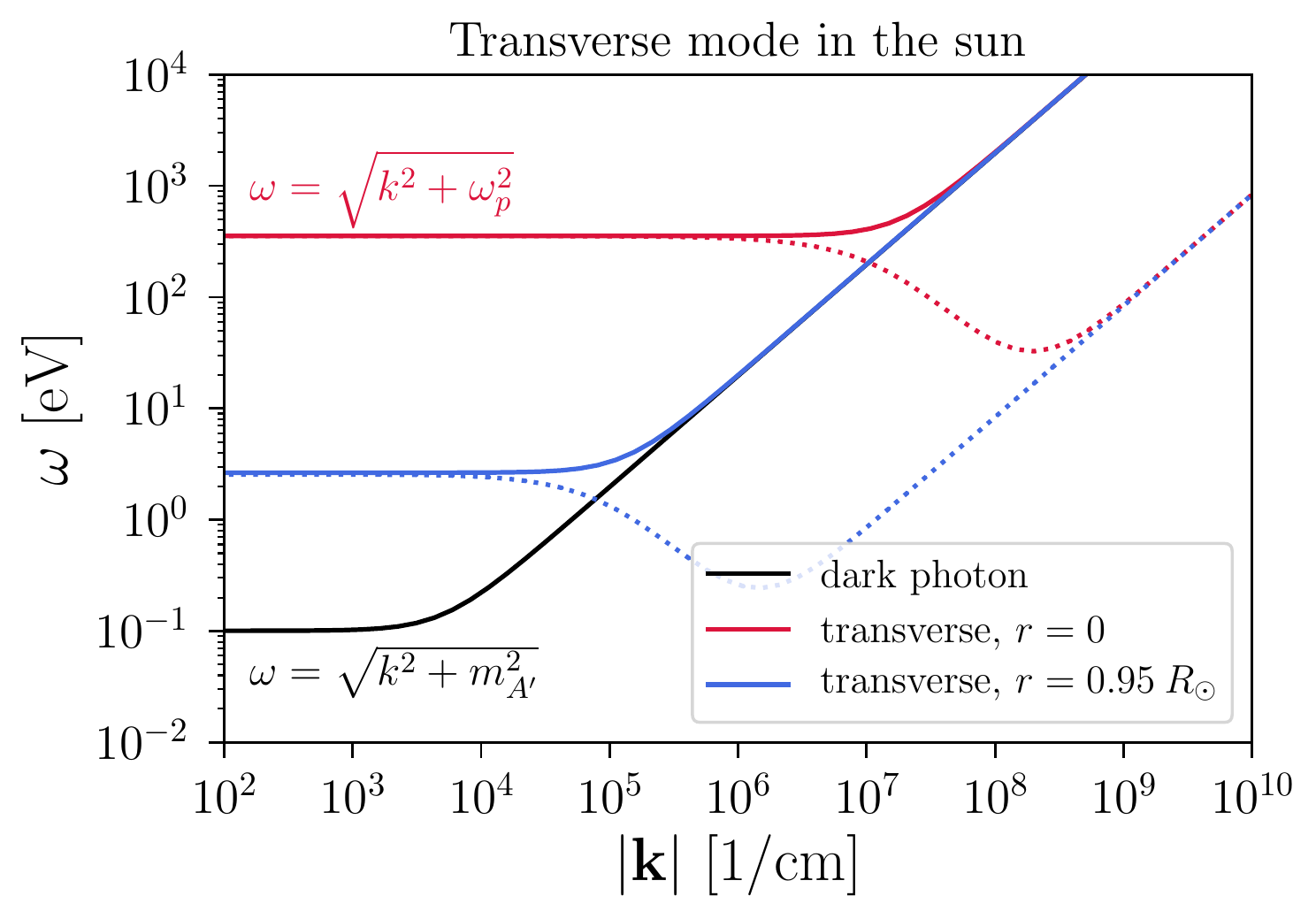}
    \caption{Dispersion relations for the transverse (\textbf{right}) and longitudinal (\textbf{left}) modes of SM photons in the sun. (\textbf{left}) The red and blue solid curves denote the dispersion relations of the longitudinal mode at the center and at the edge of the sun, respectively. The black curve shows the dispersion relation for a dark photon in a model with $m_{A^\prime}=10^{-1}\,{\rm eV}$. Dark photon resonant emission can occur when the SM photon longitudinal dispersion curve intersects the dark photon dispersion curve. This can happen for the range of $|\mathbf{k}|$ values bounded between the intersection points (circles) on the red and blue curves. We note that the longitudinal plasma wave for $|\mathbf{k}| \gtrsim 1/\lambda_{\rm D}$ in a nonrelativistic plasma suffers strong Landau damping as shown by the dashed lines. (\textbf{right}) The colored solid curves denote the dispersion relations for the transverse mode in the sun. The red (blue) dotted curve is the difference between red (blue) solid curve and the black dark photon curve for each $|\mathbf{k}|$ value. The dispersion relation curve for SM photons in the transverse mode never intersects the dispersion relation curve for a dark photon.}
    \label{fig: dispersion relation nonrelativistic}
\end{figure}

In figure~\ref{fig: dispersion relation nonrelativistic}, we take the sun as an example of the nonrelativistic plasma environment and show the dispersion relations for in-medium photons and for a dark photon. We consider a range of radius $r$ from the center of the sun to $95\%$ of the solar radius, $r\leq0.95\,R_\odot$. Electrons and protons in the sun are nonrelativistic. The dispersion relation for longitudinal EM oscillation in such an environment is $\omega \approx \omega_{\rm p}$ when $|\mathbf{k}| \lesssim 1/\lambda_{\rm D}$, where $\lambda_{\rm D}$ denotes Debye screening length. This behavior is evident for the solid lines in the left plot of figure~\ref{fig: dispersion relation nonrelativistic}. When $m_{A'}$ is less than the plasma frequency at around the edge of the sun, $\omega_{\rm p}\rvert_{r=0.95 R_\odot}\sim 1\,{\rm eV}$, the dispersion relation curve for the longitudinal EM oscillation may cross the dispersion relation curve for the dark photon at any given radius. That is, nearly the entire sun could radiate longitudinal dark photons resonantly. On the other hand, the transverse dark photon dispersion relation curve never intersects the dark photon dispersion relation curve, as is evident in the right plot of figure~\ref{fig: dispersion relation nonrelativistic}.

\section{Dark photon emission rate}\label{sec: matrix_element}

In this appendix, we calculate the lepton-pair annihilation rate and the relevant corresponding matrix elements. Here we follow the discussion of this physics in appendix~B of Ref.~\cite{Fradette:2014sza}.

The $\Gamma^{\rm prod}_{A_a}$ shown in equation~(\ref{eq: production rate}) is the annihilation rate for the processes $l\bar{l}\rightarrow A_a$ or $q\bar{q}\rightarrow A_a$. In the following, we take the lepton-pair annihilation case as an example. Denoting $p$ and $q$ as the four-momenta of two annihilating leptons, and $k=p+q$ as the four-momentum for $A_a$, the annihilation rate for the process $l\bar{l}\rightarrow A_a$ is 
\beq
\baln
    	 \Gamma^{\rm prod}_{A_a} \left(\omega\right) &= \frac{1}{2\omega}  \int \frac{d^3 {\mathbf p}}{\left(2\pi\right)^3 2E_{\mathbf p}} \frac{d^3 {\mathbf q}}{\left(2\pi\right)^3 2E_{\mathbf q}} \left(2\pi\right)^4 \delta^{\left(4\right)} \left(k-p-q\right) f_{l}\left(E_{\mathbf p}\right) f_{\bar{l}}\left(E_{\mathbf q}\right) \abs{\mathcal{M}_{l\bar{l}\rightarrow A_a}}^2\\
	 &= \frac{1}{16\pi\omega} \int \frac{{|\mathbf p|}^2 d|{\mathbf p}| \; d\cos{\theta}}{E_{\mathbf p} E_{\mathbf q}} \;  f_{l}\left(E_{\mathbf p}\right)  f_{\bar{l}}\left(E_{\mathbf q}\right)  \:  \delta\left( \omega - E_{\mathbf p} - E_{\mathbf q}  \right)  \: \abs{\mathcal{M}_{l\bar{l} \rightarrow A_a}}^2,
	 \label{eq: massive photon production rate phase space}
\ealn    
\eeq
where $\theta$ denotes the angle between the three-vectors ${\mathbf k}$ and ${\mathbf p}$. Using the identity $\delta \left( g\left(x\right) \right)  = \sum_i {\delta (x-x_i)}/{\abs{g^\prime(x_i)}}$, we write the Dirac delta function shown in equation~(\ref{eq: massive photon production rate phase space}) as
\beq
	\delta\left(  \omega - E_{\mathbf p} - E_{\mathbf q}  \right) = \sum_i {\delta\left( \cos\theta - \cos\theta_i \right)}\bigg/{ \bigg\lvert  \frac{\abs{\mathbf p} \abs{\mathbf k}}{\sqrt{ m_l^2 + \abs{\mathbf p}^2 + \abs{\mathbf k}^2 - 2 \abs{\mathbf p}\abs{\mathbf k}\cos\theta_i }} \bigg \rvert },
\eeq
where $\cos\theta_i = \frac{2 \omega E_{\mathbf p} - m_{A^\prime}{}^2}{2 \abs{\mathbf p}\abs{\mathbf k} }$. Since $\cos^2\theta_i\leq1$, we obtain the maximum and minimum values of $E_\textbf{p}$,
\beq
	E_{\rm {\bf p},max} = \frac{\omega}{2} +  \frac{\abs{\mathbf k}}{2} \sqrt{ 1 - 4\frac{m_l{}^2}{m_{A^\prime}{}^2} },~\text{and}
	\quad
	E_{\rm {\bf p},min} = \frac{\omega}{2} -  \frac{\abs{\mathbf k}}{2} \sqrt{ 1 - 4\frac{m_l{}^2}{m_{A^\prime}{}^2} }.
\label{eq: E up and low}
\eeq
(Note that $E_{\rm {\bf p},min}\geq m_e$.) Integrated over $\theta$, the annihilation rate for lepton pairs in equation~(\ref{eq: massive photon production rate phase space}) becomes 
\begin{equation}
	 \Gamma^{\rm prod}_{A_a} \left(\omega \right) = \frac{1}{16\pi\omega \abs{\mathbf k}} \int_{E_{\rm {\bf p}, min}}^{E_{\rm {\bf p}, max}} dE_{\mathbf p}  \; f_{l} \left(E_{\mathbf p}\right) f_{\bar{l}}\left(\omega - E_{\mathbf p}\right) \abs{\mathcal{M}_{l\bar{l} \rightarrow A_a}}^2.
	 \label{eq: massive photon production final}
\end{equation}
The calculation of the annihilation rate for free quarks-pairs is the same as the above calculation for the annihilation rate of lepton-pairs.

The effective coupling of $A^\prime_a$ to $J_{\rm em}^\mu$ shown in equation~(\ref{eq: in-medium mass basis}) indicates that the dark photon emission rate, $\Gamma_{A^\prime_a}^{\rm prod}$, is a factor $\kappa_{\rm eff}$ smaller than $\Gamma^{\rm prod}_{A_a}$. As a result, the \emph{integrated} dark photon emission rate in a dense medium is given by
\beq
    \bar{\Gamma}_{A^\prime_a}^{\rm prod} = \int \frac{d^3\mathbf{k}}{\left(2\pi\right)^3} \; \kappa_{{\rm eff},a}^2 \: \Gamma^{\rm prod}_{A_a}\left(\omega\right).
\eeq

The matrix element in equation~(\ref{eq: massive photon production final}) is  
\beq
    \mathcal{M}_{l\bar{l}\rightarrow A_a} = \bar{v}\left(p_l\right) \left(-ie\gamma^\mu \right) u\left(q_{\bar{l}}\right) \varepsilon_{a,\mu}^{*}  \left(k\right),
\eeq
where $\varepsilon_{a}^{\mu}$ is the external polarization state of photon $A^{\mu}$. The squared matrix element for the transverse mode is evaluated by summing over the initial lepton spin states and the two final transverse photon states. This gives
\beq
\baln
    \sum_{l\bar{l}~{\rm spins},\:\rm \pm{\rm T}}\abs{\mathcal{M}_{l\bar{l} \rightarrow A_{\pm{\rm T}}}}^2 
    &= 4e^2 \: {\rm Tr}\bigg[\left(\slashed{q}-m\right)\gamma^\mu\left(\slashed{p}+m\right)\gamma^\nu\bigg] \left(0,1,1,0\right)_{\rm\mu\nu,\:diag}\\
    &= 16\pi \alpha\left(m_{A^\prime}{}^2-2|\mathbf{p}|^2\sin^2\theta\right).
\ealn    
\eeq    
The squared matrix element for longitudinal mode is evaluated by summing over initial lepton spin states. This gives
\beq
\baln
    \sum_{l\bar{l}~{\rm spins}}\abs{\mathcal{M}_{l\bar{l} \rightarrow A_{\rm L}}}^2 
    &= 4e^2 \: {\rm Tr}\bigg[\left(\slashed{q}-m\right)\gamma^\mu\left(\slashed{p}+m\right)\gamma^\nu\bigg] 
    \left(\begin{array}{cccc}
         k^2 & 0 & 0 & -\omega k \\
         0 & 0 & 0 & 0 \\
         0 & 0 & 0 & 0 \\
         -\omega k & 0 & 0 & \omega^2 \\
        \end{array}\right)_{\mu\nu}\\
    &= 16\pi\alpha \bigg[ \frac{1}{2} m_{A^\prime}{}^2 - \frac{2}{m_{A^\prime}{}^2} \left( \abs{\mathbf k} E_{\mathbf p} - \omega \abs{\mathbf p} \cos\theta \right)^2 \bigg].
\ealn    
\eeq

\bibliographystyle{JHEP}
\bibliography{DP_BBN}

\end{document}